\documentstyle[12pt,epsf,a4,epsfig]{article}

\newcommand{\beq}{\begin{equation}}
\newcommand{\eeq}{\end{equation}}
\newcommand{\beqa}{\begin{eqnarray}}
\newcommand{\eeqa}{\end{eqnarray}}
\newcommand{\no}{\nonumber}
\newcommand{\q}{\quad}
\newcommand{\qq}{\qquad}

\begin{document}

\hfill 

\hfill 

\bigskip\bigskip

\begin{center}

{{\Large\bf Threshold $\eta$ and $\eta'$ electroproduction off nucleons
\footnote{Work supported in part by the DFG}}}

\end{center}

\vspace{.4in}

\begin{center}
{\large B. Borasoy\footnote{email: borasoy@physik.tu-muenchen.de}}

\bigskip

\bigskip

Physik Department\\
Technische Universit{\"a}t M{\"u}nchen\\
D-85747 Garching, Germany \\

\vspace{.2in}

\end{center}

\vspace{.7in}

\thispagestyle{empty} 

\begin{abstract}
The electroproduction of $\eta$ and $\eta'$ mesons on the proton and the
neutron is investigated at tree level within the framework of $U(3)$ 
chiral perturbation theory. In addition to the Born terms 
low-lying resonances such as
the vector mesons and $J^P= 1/2^+, 1/2^-$ baryon resonances are included
explicitly and their contributions are calculated.
Results for the separated differential cross sections are
presented. 
\end{abstract}

\vfill

\section{Introduction}  
In electroproduction one can get detailed information about the structure
of the nucleon due to the longitudinal coupling of the virtual photon to the
nucleon spin. It furthermore is a tool to study baryon resonances
and the investigation of transitions between these states provides a crucial 
test for hadron models.
Perturbative QCD should apply at sufficiently high photon virtuality $|k^2|$,
see e.g. \cite{CP, LDW}, however there is no consensus how high the momentum
transfer must be. It has been found experimentally, that in the case of
electroproduction of the $\Delta (1232)$ resonance at momentum transfers up to
$|k^2| = 4.0$ GeV$^2$, perturbative QCD is not applicable, \cite{eldel}, 
whereas a possible onset of scaling in the reaction 
$ e +p \rightarrow e+p + \eta$ at $|k^2| = 3.6$ GeV$^2$ is reported in
\cite{els11}. 
At low $|k^2|$ non-perturbative QCD dominates. This region has been treated,
e.g., by incorporating relativistic effects into the constituent quark model
\cite{IK}, using light-front approaches \cite{CK} and within the context of
chiral perturbation theory \cite{BKM}.

Because of their hadronic decay modes nucleon resonances have large overlapping
widths, which makes it difficult to study individual states, but selection
rules in certain decay channels can reduce the number of possible resonances.
The isoscalars $\eta$ and $\eta'$ are such examples since, due to isospin
conservation, only the isospin-$\frac{1}{2}$ excited states decay into the
$\eta N$ and $\eta' N$ channels.
In recent years both $\eta$ and $\eta'$ electro- and photoproduction 
have been of
considerable interest. The $\eta$ photoproduction of protons has been measured
at MAMI \cite{Kru} and resonance parameters of the $S_{11} (1535) $
resonance and the
electromagnetic coupling $\gamma p \rightarrow  S_{11}$ 
have been extracted from the data.
Photoproduction of the $\eta'$ has been measured at ELSA \cite{Pl}.
The experimental data for electroproduction of the $\eta$ is still very scarce;
it is limited to a few older Bonn data \cite{Br} and recently published data
from CEBAF \cite{els11} at high momentum transfer.

On the theoretical side, the $\eta$ meson has been treated as a pure $SU(3)$
octet state $\eta_8$ and mixing of $\eta_8$ with
the corresponding singlet state  $\eta_0$ which yields the physical states
$\eta$ and  $\eta'$ is generally neglected.
The  $\eta'$ is interesting by itself. The QCD Lagrangian with massless quarks
exhibits an $SU(3)_L \times SU(3)_R$ chiral symmetry which is broken down
spontaneously to $SU(3)_V$, 
giving rise to a Goldstone boson octet of pseudoscalar mesons
which become massless in the chiral limit of zero quark masses.
On the other hand, the axial $U(1)$ symmetry of the QCD Lagrangian is broken by
the anomaly. 
The corresponding pseudoscalar singlet would otherwise have a mass comparable
to the pion mass \cite{W}. Such a particle is missing in the spectrum and the
lightest candidate would be the $\eta'$ with a mass of 958 MeV which is
considerably heavier than the octet states.
In conventional chiral perturbation theory the $\eta'$ 
is not included explicitly, although it does show up in the form of a
contribution to a coupling coefficient of the Lagrangian, a so-called
low-energy constant (LEC).
Recently, the $\eta'$ has been included in baryon chiral perturbation theory 
in a systematic fashion \cite{B}. Using this approach $\eta$ and $\eta'$
photoproduction off the nucleons has been investigated \cite{B2}.
Low-lying resonances such as
the vector mesons and $J^P= 1/2^+, 1/2^-$ baryon resonances are included
explicitly and their contributions together with the Born terms are
calculated. The coupling constants of the resonances are determined from strong
and radiative decays and reasonable agreement with experimental data
near threshold is obtained.

The purpose of this paper is to extend this approach to the electroproduction
of $\eta$ and $\eta'$ on the nucleons which provides a further test for this
simple model. From such a simplified treatment of $\eta$ and $\eta'$
electroproduction one should not expect to forecast experimental data in
detail; here we are rather concerned with qualitative agreement.
In order to obtain
a better description of the experimental data, one has to include chiral loops
and further resonances, but this is beyond the scope of the present
investigation.
This work should therefore be considered to be mainly a check if the inclusion
of $\eta$ and $\eta'$ mesons in a nonet of pseudoscalar mesons as proposed in
\cite{B} leads to an adequate description for processes of $\eta$
and $\eta'$  mesons with baryons.

In the next section we present the necessary formalism for  electroproduction
of $\eta$ and $\eta'$ mesons. The effective chiral Lagrangian including
explicitly low-lying resonances is given in Sec. 3. The invariant amplitudes
are shown in Sec. 4 together with the numerical results. We conclude with a
summary in Sec. 5.

\section{General Formalism}
The $T$-matrix element for the processes $N(p_1) + \gamma^*(k) \rightarrow
N(p_2) + \phi(q)$ with $\phi = \eta$ or $\eta^{\prime}$ is given by
\beq
\langle  p_2, q \; \mbox{out} | p_1, k \; \mbox{in} \rangle  =
\delta_{fi} + ( 2 \pi)^4 i \delta^{(4)}(p_2 + q - p_1 - k) T_{fi} .
\eeq
The Mandelstam variables are
\beqa
s &=& (k + p_1)^2 = (q + p_2)^2 \no \\
t &=& (k -q)^2 = (p_1 - p_2)^2 \no \\
u &=& (k - p_2)^2 = (q - p_1)^2
\eeqa
subject to the constraint $ s+t+u= 2 M_N^2 + m_\phi^2 + k^2$ with $M_N$ and
$m_\phi$ being the mass of the nucleon and the pseudoscalar meson,
respectively. The invariant four-momentum transfer squared, $t$, can be related
to the scattering angle $\vartheta$ in the c.m. system via
\beq
t = m^2_\phi - 2 q^0 k^0 + 2 |{\bf q}| |{\bf k}| z + k^2
\eeq
with $z = \cos \vartheta$.

In general, $T$ can be decomposed as
\beq
T_{fi} = i \epsilon_\mu \bar{u}_2 \sum_{i =1}^{8} B_i {\cal N}_i^\mu u_1
\eeq
with the invariant amplitudes
\beqa
{\cal N}^1_\mu & =& \gamma_5 \gamma_\mu k \! \! \! / ,  \qq 
{\cal N}^2_\mu = 2 \gamma_5 P_\mu , \qq 
{\cal N}^3_\mu = 2 \gamma_5 q_\mu , \qq
{\cal N}^4_\mu = 2 \gamma_5 k_\mu ,\no \\
{\cal N}^5_\mu & =& \gamma_5 \gamma_\mu,  \qq 
{\cal N}^6_\mu = \gamma_5 k \! \! \! /  P_\mu , \qq 
{\cal N}^7_\mu = \gamma_5 k \! \! \! /  k_\mu , \qq
{\cal N}^8_\mu = \gamma_5 k \! \! \! /  q_\mu  
\eeqa
and $P = \frac{1}{2}( p_1 + p_2)$.
From current conservation one obtains the relations
\beqa
k^2 B_1 + k \cdot ( p_1 + p_2) B_2 + 2 k \cdot q B_3 + 2 k^2 B_4 = 0 \no \\
B_5 + \frac{1}{2} k \cdot ( p_1 + p_2) B_6 + k^2 B_7 +  k \cdot q B_8 = 0
\eeqa
which are used to eliminate $B_3$ and $B_5$.
It is therefore more convenient to define a set of independent amplitudes
\beq
T_{fi} = i \bar{u}_2 \sum_{i =1}^{6} A_i {\cal M}_i u_1
\eeq
with
\beqa
{\cal M}_1 &=& \frac{1}{2} \gamma_5 \gamma_\mu \gamma_\nu F^{\mu \nu}, \qq
{\cal M}_2 = 2 \gamma_5 P_\mu (q- \frac{1}{2}k)_\nu F^{\mu \nu}, \no \\
{\cal M}_3 &=& \gamma_5 \gamma_\mu q_\nu F^{\mu \nu}, \qq
{\cal M}_4  = 2 \gamma_5 \gamma_\mu P_\nu F^{\mu \nu} - 2M_N {\cal M}_1  , \no
\\ 
{\cal M}_5 &=& \gamma_5 k_\mu q_\nu F^{\mu \nu}, \qq
{\cal M}_6 = \gamma_5 k_\mu \gamma_\nu F^{\mu \nu}
\eeqa
and $F_{\mu \nu} = \epsilon_\mu k_\nu - \epsilon_\nu k_\mu$.
The $A_i$ obey the crossing relations
\beqa
A_i(s,u) &=&  A_i(u,s)   \qq  i = 1,2,4 \no \\
A_i(s,u) &=&  -A_i(u,s)   \qq  i = 3,5,6
\eeqa
and are related to the $B_i$ via
\beqa
A_1 &=& B_1 - M_N B_6 , \qq \q
A_2 = \frac{2}{m_\phi^2 -t} \, B_2 , \qq \q
A_3 = -B_8 , \no \\
A_4 &=& - \frac{1}{2} B_6 , \qq \q
A_5 = \frac{2}{s+u - 2 M_N^2} \Big( B_1 - \frac{s-u}{2 ( m_\phi^2 -t)} B_2 + 
       2 B_4 \Big)  , \no \\
A_6 &=& B_7 .
\eeqa
The unpolarized $\eta$/$\eta'$ electroproduction triple differential cross
section reads
\beq
\frac{d \sigma}{d E_f d\Omega_f d\Omega_\phi} =
\frac{\alpha E_f ( s - M_N^2)}{4 \pi^2 E_i M_N k^2 ( \epsilon -1)} 
\frac{d \sigma}{d\Omega_\phi}
\eeq
with $\alpha = e^2/4\pi$ the fine structure constant and $E_{i/f}$ is the
laboratory energy of the incoming/outgoing electron.
The photon polarization $\epsilon$ is given by
\beq
\epsilon^{-1} = 1 + 2 \Big(1- \frac{k_0^2}{k^2} \Big) \tan^2 \frac{\psi}{2}
\eeq
with $k_0$ the photon energy and $\psi$ the photon scattering angle.
The differential cross section can be decomposed into transverse ($T$),
longitudinal ($L$), transverse-longitudinal ($TL$) and transverse-transverse
($TT$) pieces
\beq
\frac{d \sigma}{d\Omega_\phi} = \frac{2 \sqrt{s} |{\bf q}|}{s-M_N^2}
\Big( R_T + \epsilon_L R_L + \sqrt{ 2 \epsilon_L ( 1+ \epsilon)} \cos \phi
R_{TL} + \epsilon \cos 2 \phi R_{TT} \Big) ,
\eeq
where the $R_I (I=T,L,TL,TT)$ are called structure functions, $\phi$ is the
azimuthal angle between the scattering and the reaction plane, and $\epsilon_L$
is the longitudinal photon polarization
\beq
\epsilon_L = - \frac{k^2}{k_0^2} \epsilon .
\eeq
The separated virtual photon cross sections are
\beq
\frac{d \sigma_I}{d\Omega_\phi} = \frac{2 \sqrt{s} |{\bf q}|}{s-M_N^2} R_I,
   \qq \q I = T,L,TL,TT .
\eeq
Since we restrict ourselves to the threshold region, it is convenient to
perform a multipole decomposition and confine ourselves to $S$- and $P$-waves.
To this end, one expresses the transition amplitude in terms of Pauli spinors
and matrices
\beq
\frac{1}{8 \pi \sqrt{s}} i \bar{u}_2 \sum_{i =1}^{6} A_i {\cal M}_i u_1=
\chi^\dagger_2 {\cal F} \chi_1 .
\eeq
The matrix ${\cal F}$ can be written as
\beqa
{\cal F} &=& i \mbox{\boldmath$ \sigma$} \cdot 
      {\bf b} {\cal F}_1 +
   \mbox{\boldmath$\sigma$} \cdot {\bf \hat{q}} \, 
     \mbox{\boldmath$ \sigma$} \cdot ( {\bf \hat{k}} \times
   {\bf b} ) \, {\cal F}_2 + i \mbox{\boldmath$ \sigma$} 
    \cdot {\bf \hat{k}} \,
   {\bf \hat{q}} \cdot {\bf b} \, {\cal F}_3  \no \\
&& +    i \mbox{\boldmath$ \sigma$} \cdot {\bf \hat{q}} \, {\bf \hat{q}} 
     \cdot {\bf b} \, {\cal F}_4  - 
 i \mbox{\boldmath$\sigma$} \cdot {\bf \hat{q}} \, b_0 {\cal F}_7 
 - i \mbox{\boldmath$ \sigma$}  \cdot {\bf \hat{k}} \,b_0 {\cal F}_8 ,
\eeqa
where
\beq
b_\mu = \epsilon_\mu - \frac{1}{|{\bf k|}} \mbox{\boldmath$ \epsilon$} 
\cdot {\bf \hat{k}} \, k_\mu .
\eeq
With this choice of gauge, the virtual photon has only scalar and transverse
components. The ${\cal F}_i$ are related to the $A_i$ via
\beqa
{\cal F}_1  &=&  (\sqrt{s} - M_N) \, \frac{N_1 N_2}{8 \pi \sqrt{s}}\no \\
&&  \times \Big[ A_1 + \frac{k \cdot q}{\sqrt{s} - M_N} A_3 + \Big(
      \sqrt{s} - M_N - \frac{k \cdot q}{\sqrt{s} - M_N} \Big) A_4 
   - \frac{k^2}{ \sqrt{s} - M_N} A_6  \Big]  \no \\
{\cal F}_2  &=&  (\sqrt{s} + M_N) \, \frac{N_1 N_2}{8 \pi \sqrt{s}} \, 
  \frac{|{\bf q}| |{\bf k}|}{(E_1 + M_N)(E_2 + M_N)} \no \\
&&  \times \Big[-A_1 + \frac{k \cdot q}{\sqrt{s} + M_N} 
 A_3 + \Big(\sqrt{s} + M_N - \frac{k \cdot q}{\sqrt{s} + M_N} \Big) A_4 
      - \frac{k^2}{ \sqrt{s} + M_N} A_6 \Big]  \no \\
{\cal F}_3  &=&  (\sqrt{s} + M_N) \, \frac{N_1 N_2}{8 \pi \sqrt{s}} \, 
   \frac{|{\bf q}| |{\bf k}|}{E_1 + M_N}   \no \\
&&  \times \Big[  \frac{M_N^2 -s +
   \frac{1}{2} k^2}{\sqrt{s} + M_N} A_2 + A_3 - A_4 - 
   \frac{k^2}{ \sqrt{s} + M_N} A_5 \Big]  \no \\
 {\cal F}_4  &=&  (\sqrt{s} - M_N) \, \frac{N_1 N_2}{8 \pi \sqrt{s}}\, 
  \frac{|{\bf q}|^2}{E_2 + M_N}  \no \\
&&  \times \Big[  \frac{s -M_N^2 -
   \frac{1}{2} k^2}{\sqrt{s} - M_N} A_2 + A_3 - A_4 + 
   \frac{k^2}{ \sqrt{s} - M_N} A_5 \Big] \no \\
 {\cal F}_7  &=&  \frac{N_1 N_2}{8 \pi \sqrt{s}}\, 
   \frac{|{\bf q}|}{E_2 + M_N}\, \bigg[ - (E_1 -M_N) A_1 \no \\
& & - \frac{1}{2 k_0} 
   \Big( |{\bf k}|^2 (2 k_0 \sqrt{s} - 3 k \cdot q) - {\bf q} \cdot {\bf k} (2
   s - 2 M_N^2 - k^2) \Big) A_2 \no \\
&& +  ( q_0 (\sqrt{s} -M_N) - k \cdot q) A_3 \no \\
&& + ( k \cdot q - q_0 (\sqrt{s}
-M_N)  + (E_1 -M_N) (\sqrt{s} + M_N)) A_4 \no \\
&& 
 + ( q_0 k^2 - k_0 k \cdot q) A_5 - (E_1 -M_N) (\sqrt{s} + M_N) A_6 \bigg]
\no \\
 {\cal F}_8  &=&  \frac{N_1 N_2}{8 \pi \sqrt{s}}\, 
   \frac{|{\bf k}|}{E_1 + M_N}\, \bigg[  (E_1 +M_N) A_1 \no \\
&&+  \frac{1}{2 k_0} 
   \Big( |{\bf k}|^2 (2 k_0 \sqrt{s} - 3 k \cdot q) - {\bf q} \cdot {\bf k} (2
   s - 2 M_N^2 - k^2) \Big) A_2 \no \\
&& +  ( q_0 (\sqrt{s} +M_N) - k \cdot q) A_3 \no \\
&&+ ( k \cdot q - q_0 (\sqrt{s}
+M_N)  + (E_1 +M_N) (\sqrt{s} - M_N)) A_4 \no \\
&&
- ( q_0 k^2 - k_0 k \cdot q) A_5 - (E_1 +M_N) (\sqrt{s} - M_N) A_6 \bigg]
\eeqa
with
\beq
N_i = \sqrt{M_N + E_i}, \qq  E_i = \sqrt{M_N^2 + {\bf p}_i^2} .
\eeq
The projection matrices for the lowest multipoles $E_{0+}, M_{1+}, M_{1-},E_{1+},
L_{0+},L_{1+}$ and $L_{1-}$ are given by
\beq
\left( \begin{array}{l}  E_{0+} \\[0.1cm]  M_{1+} \\[0.1cm]  M_{1-} \\[0.1cm]
   E_{1+} \end{array}
\right) =   \int_{-1}^1 d z
\left( \begin{array}{cccc}
\frac{1}{2} P_0 & - \frac{1}{2} P_1 & 0 & \frac{1}{6} [P_0 - P_2] \\[0.1cm]
\frac{1}{4} P_1 & - \frac{1}{4} P_2 & \frac{1}{12} [P_2 - P_0] & 0 \\[0.1cm]
- \frac{1}{2} P_1 & \frac{1}{2} P_0 & \frac{1}{6} [P_0 - P_2] & 0 \\[0.1cm]
\frac{1}{4} P_1 & - \frac{1}{4} P_2 & \frac{1}{12} [P_0 - P_2] & 
\frac{1}{10} [P_1 - P_3] \end{array}
\right)
\left( \begin{array}{l}  {\cal F}_1 \\[0.1cm] {\cal F}_2 \\[0.1cm]
    {\cal F}_3 \\[0.1cm]  {\cal F}_4 
\end{array} \right)
\eeq
and
\beq
\left( \begin{array}{l}  L_{0+} \\[0.1cm]  L_{1+} \\[0.1cm]  L_{1-} \end{array}
\right) =   \frac{k_0}{|{\bf k}|} \int_{-1}^1 d z
\left( \begin{array}{cc}
\frac{1}{2} P_1 & \frac{1}{2} P_0 \\[0.1cm]
\frac{1}{4} P_2 & \frac{1}{4} P_1 \\[0.1cm]
\frac{1}{2} P_0 & \frac{1}{2} P_1 \end{array}
\right)
\left( \begin{array}{l}  {\cal F}_7 \\[0.1cm] {\cal F}_8 \end{array} \right)
\eeq
with $P_i$ being the Legendre polynomials. The scalar multipoles are related to
the longitudinal multipoles by $S_{l+} = (|{\bf k}|/k_0) L_{l+}$.
The structure functions $R_I$ can be expressed in terms of the multipoles as
follows
\beqa
R_T &=& |E_{0+} + \cos \vartheta P_1|^2 + \frac{1}{2} \sin^2 \vartheta \,
   (|P_2|^2 + |P_3|^2 ) \no \\
R_L &=& |L_{0+} + \cos \vartheta P_4|^2 + \sin^2 \vartheta \, |P_5|^2 \no \\
R_{TL} &=& - \sin \vartheta \,
   \mbox{Re} \Big[ (E_{0+} + \cos \vartheta P_1) P_5^*
  + (L_{0+} + \cos \vartheta P_4) P_2^* \Big] \no \\
R_{TT} &=& \frac{1}{2} \sin^2 \vartheta \, ( |P_2|^2 - |P_3|^2 )
\eeqa
with the combinations
\beqa
P_1 &=& 3 E_{1+} + M_{1+} - M_{1-} , \qq P_2 = 3 E_{1+} - M_{1+} + M_{1-} , 
\no \\
P_3 &=& 2 M_{1+}+ M_{1-} ,  \qq P_4 = 4 L_{1+} + L_{1-} , \qq 
P_5 = L_{1-} - 2 L_{1+}.
\eeqa
This completes the necessary formalism needed in the present investigation.

\section{The effective Lagrangian}
In this section, we will introduce the effective Lagrangian with $\eta$ and
$\eta'$ coupled both to the ground state baryon octet and low-lying
resonances in the $s,u$- and $t$-channel. A systematic framework  for
the $\eta'$ in baryon chiral perturbation theory has been developed in
\cite{B} and extended by including explicitly low-lying meson and baryon
resonances \cite{B2}.
Our starting point is the $U(3)_L \times U(3)_R$ chiral effective Lagrangian of
the pseudoscalar meson nonet $(\pi,K,\eta_8,\eta_0)$ 
coupled to the ground state
baryon octet $(N,\Lambda, \Sigma, \Xi)$ at lowest order in the derivative
expansion
\beq
{\cal L} = {\cal L}_\phi + {\cal L}_{\phi B}
\eeq
with
\beq  \label{mes}
{\cal L}_\phi = 
- v_0 \eta_0^2 + \frac{F_\pi^2}{4}  \langle u_{\mu} u^{\mu} \rangle 
+ \frac{F_\pi^2}{4} \langle \chi_+ \rangle + 
   i F_0 v_3 \eta_0 \langle \chi_- \rangle
+ \frac{1}{12} (F_0^2 - F_\pi^2)  
   \langle u_{\mu} \rangle \langle u^{\mu} \rangle 
\eeq
and
\beqa  \label{bar}
{\cal L}_{\phi B} &=& i \langle \bar{B}  \gamma_{\mu}  [D^{\mu},B] \rangle 
 - M_N \langle \bar{B}B \rangle - \frac{1}{2} D \langle \bar{B} \gamma_{\mu}
 \gamma_5 \{u^{\mu},B\} \rangle  \no \\
&& - \frac{1}{2} F \langle \bar{B} \gamma_{\mu} \gamma_5 [u^{\mu},B] \rangle 
- \lambda \langle \bar{B} \gamma_{\mu} \gamma_5 B \rangle 
  \langle u^{\mu} \rangle \no \\
&& + \frac{1}{8 M_N} b_6^D
   \langle \bar{B} \sigma^{\mu \nu} \{ F^+_{\mu \nu}, B \} \rangle 
+  \frac{1}{8 M_N} [1+b_6^F ]
   \langle \bar{B} \sigma^{\mu \nu} [ F^+_{\mu \nu}, B ] \rangle .
\eeqa
The pseudoscalar meson nonet is summarized in a matrix valued 
field $U(x)$
\begin{equation}
 U(\phi,\eta_0) = u^2 (\phi,\eta_0) = 
\exp \lbrace 2 i \phi / F_\pi + i \sqrt{\frac{2}{3}} \eta_0/ F_0 \rbrace  
, 
\end{equation}
where $F_\pi \simeq 92.4$ MeV is the pion decay constant and the singlet
$\eta_0$ couples to the singlet axial current with strength $F_0$.
The unimodular part of the field $U(x)$ contains the degrees of freedom of
the Goldstone boson octet $\phi$
\begin{eqnarray}
 \phi =  \frac{1}{\sqrt{2}}  \left(
\matrix { {1\over \sqrt 2} \pi^0 + {1 \over \sqrt 6} \eta_8
&\pi^+ &K^+ \nonumber \\
\pi^-
        & -{1\over \sqrt 2} \pi^0 + {1 \over \sqrt 6} \eta_8 & K^0
        \nonumber \\
K^-
        &  \bar{K^0}&- {2 \over \sqrt 6} \eta_8  \nonumber \\} 
\!\!\!\!\!\!\!\!\!\!\!\!\!\!\! \right) \, \, \, \, \, ,  
\end{eqnarray}
while the phase det$U(x)=e^{i\sqrt{6}\eta_0/F_0}$
describes the $\eta_0$.\footnote{For details the reader is referred to
\cite{B}.} 
In order to incorporate the baryons into the effective theory it is convenient
to form an object of axial-vector type with one derivative
\beq
u_{\mu} = i u^\dagger \nabla_{\mu} U u^\dagger
\eeq
with $\nabla_{\mu}$ being the covariant derivative of $U$.
The expression $\langle \ldots \rangle$ denotes the trace in flavor space
and the quark mass matrix ${\cal M} = \mbox{diag}(m_u,m_d,m_s)$
enters in the combinations
\beq
\chi_\pm = 2 B_0 ( u {\cal M} u \pm  u^\dagger {\cal M} u^\dagger)
\eeq
with $B_0 = - \langle  0 | \bar{q} q | 0\rangle/ F_\pi^2$ the order
parameter of the spontaneous symmetry violation.
Expanding the Lagrangian ${\cal L}_\phi$ in terms of the meson fields
one observes terms quadratic in the meson fields that contain the factor
$\eta_0 \eta_8$ which leads to $\eta_0$-$\eta_8$ mixing.
Such terms arise from the explicitly symmetry breaking terms 
$\frac{F_\pi^2}{4} \langle \chi_+ \rangle + i F_0 v_3 \eta_0 
\langle \chi_- \rangle$ and read
\beq
- \Big( \frac{2 \sqrt{2}}{3} \frac{F_\pi}{F_0} + \frac{8}{\sqrt{3}} 
\frac{F_0}{F_\pi} v_3 \Big) B_0 (\hat{m} -m_s) \eta_0 \eta_8
\eeq
with $\hat{m} = \frac{1}{2} (m_u + m_d)$.
The states $\eta_0$ and $\eta_8$ are therefore not mass eigenstates.
The mixing yields the eigenstates $\eta$ and $\eta'$,
\beqa
| \eta \rangle & = & \cos \theta \, | \eta_8 \rangle -
                     \sin \theta \, | \eta_0 \rangle  \no \\
| \eta'\rangle & = & \sin \theta \, | \eta_8 \rangle +
                     \cos \theta \, | \eta_0 \rangle , 
\eeqa
which is valid in the leading order of flavor symmetry breaking and
we have neglected other pseudoscalar isoscalar states which could mix
with both $\eta_0$ and $\eta_8$.
The $\eta$-$\eta'$ mixing angle can be determined from the two
photon decays of $\pi^0, \eta, \eta'$, which require a mixing angle
around -20$^\circ$ \cite{exp}. 
We will make use of this experimental input in order to
diagonalize the mass terms of the effective mesonic Lagrangian.

The baryonic Lagrangian consists of the free kinetic term and the axial-vector
couplings of the mesons to the baryons. The values of the LECs $D$ and $F$ can
be extracted from semileptonic hyperon decays.
A fit to the experimental data delivers 
$D=0.80 \pm 0.01$ and $F=0.46 \pm 0.01$ \cite{CR}.
From a fit to $\eta$ and $\eta'$ photoproduction one obtains for the
axial flavor-singlet coupling $\lambda = 0.05$ \cite{B2}.
The covariant derivative of the baryon field is given by
\beq
[ D_\mu, B] = \partial_\mu B + [ \Gamma_\mu, B]
\eeq
with the chiral connection
\beq
\Gamma_\mu \simeq  - i v_\mu =  i e Q {\cal A}_\mu
\eeq
to the order we are working and $Q = \frac{1}{3} \mbox{diag}(2,-1,-1)$
is the quark charge matrix.
Note that there is no pseudoscalar coupling of $\eta_0$ to the baryons of
the form $\eta_0 \bar{B}  \gamma_5 B$. Such a term is in principle possible but
can be absorbed by the $\lambda$-term in Eq. (\ref{bar}) 
by means of the equation of motion for the baryons. 
We also take the magnetic moments of the baryons at leading order into
account which are given by the two terms $b_6^D$ and $b_6^F$.
The quantity $F_{\mu \nu}^+$ contains the electromagnetic field strength
tensor $F_{\mu \nu}$ of the external vector field $v_\mu$
\beqa
F_{\mu \nu}^+  &\equiv&   u^\dagger F_{\mu \nu} u  +
          u F_{\mu \nu} u^\dagger  \no \\
  &=&   2 ( \partial_\mu v_\nu - \partial_\nu v_\mu ) + {\cal
  O}(\phi^2)  .
\eeqa
Although these terms are of higher chiral order, 
they might lead to some substantial
contributions and will therefore be included in this investigation.
A detailed analysis of the baryon magnetic moments in chiral perturbation
theory has been given in \cite{SM}. A least-squares fit to the magnetic moments
of the baryon octet leads to $b_6^D=2.39$ and $b_6^F=0.77$.

We now proceed by including explicitly low-lying resonances in our theory.
In the $t$-channel the lowest-lying resonances are the octet of the
vector mesons $(\rho, K^*, \omega)$. Note  that $\phi$
exchange has been found to be almost negligible \cite{B2} which
is in agreement with the OZI suppression.
The coupling of the baryons to the vector mesons is given by
\beq  \label{vnn}
{\cal L}_{VBB} = \frac{1}{2} \bar{p} \Gamma_\mu p  \Big(
   g_{\rho N} \rho_0^\mu + g_{\omega N} \omega^\mu \Big)
+  \frac{1}{2} \bar{n} \Gamma_\mu n  \Big(
   - g_{\rho N} \rho_0^\mu + g_{\omega N} \omega^\mu \Big),
\eeq
where the operator $\Gamma_{\mu}$ involves a vector and a tensor coupling
\beq
\Gamma_{\mu} = \gamma_{\mu} + i \frac{\kappa_V}{2 M_N} \sigma_{\mu \nu}
                 ( p'-p)^\nu .
\eeq
The couplings $g_{VN}$ are quite well known, we use $g_{\rho N} = 6.08$ and 
$g_{\omega N} = 3 g_{\rho N}$ \cite{Ba}.
Furthermore, the tensor coupling for the $\rho$ meson is
given by $\kappa_\rho =6$, whereas $\kappa_\omega \simeq 0$. Instead of using a
common tensor coupling $\kappa_V$ for $\rho$ and $\omega$, as prescribed by the
Lagrangian in Eq. (\ref{vnn}), we prefer to work with the physical values
$\kappa_\rho =6$ and $\kappa_\omega = 0$.

The electromagnetic piece of the Lagrangian is given by
\beq  \label{elec}
{\cal L}_{V \gamma \phi} =  e \, \epsilon^{\mu \nu \alpha \beta}
    \partial_\alpha {\cal A}_\beta \Big( \partial_\nu \eta 
  \sum_{V= \rho_0, \omega,\phi} g_{V \gamma \eta} V_\mu
 +  \partial_\nu \eta' \sum_{V= \rho_0, \omega,\phi} g_{V \gamma \eta'} V_\mu
 \Big).
\eeq
The experimental values for the $g_{V \gamma \eta}$ can be extracted from the
decay width of radiative decays of the vector mesons
\beq
\Gamma( V \rightarrow \eta \gamma) = \frac{e^2}{96 \pi} g_{V \gamma \eta}^2
    ( m_V - \frac{m_\eta^2}{m_V} )^3 .
\eeq
Using the values for $\Gamma( \rho \rightarrow \eta \gamma)$ and  
$\Gamma( \omega \rightarrow \eta \gamma)$ from \cite{exp} we obtain
\beqa   \label{gvge}
g_{ \rho \gamma \eta}  &=&  1.46 \pm 0.16  \; \mbox{GeV}^{-1} \no \\ 
g_{ \omega \gamma \eta}  &=&  0.53 \pm 0.04   \; \mbox{GeV}^{-1}  ,
\eeqa
where the uncertaimty in the couplings stems from the given experimental
errors. 
The coupling strength of the vector mesons to the $\eta'$
can be extracted directly from the
decay widths of the pertinent radiative $\eta'$ decays 
\beq
\Gamma( \eta' \rightarrow V \gamma) = \frac{e^2}{32 \pi} g_{V \gamma \eta'}^2
    ( m_{\eta'} - \frac{m_V^2}{m_{\eta'}} )^3 .
\eeq
We obtain
\beqa   \label{gvgep}
g_{ \rho \gamma \eta'}  &=&  1.31  \pm 0.04 \; \mbox{GeV}^{-1} \no \\ 
g_{ \omega \gamma \eta'}  &=&  0.45  \pm 0.04 \; \mbox{GeV}^{-1}  .
\eeqa
This determines completely the contributions of the vector mesons.
Note that the vector meson contribution is usually reduced, e.g., 
by using a form factor \cite{ZMB}. However,
this effect should be reasonably small for $\eta$ and $\eta'$ 
photo- and electroproduction close to threshold.

Baryon resonances contribute in the $s$- and $u$-channel. In this work we
consider the lowest-lying $S$- and $P$-wave baryon resonances, i.e. the $J^P =
1/2^+$ and $1/2^-$ octets which include $P_{11}(1440)$ and
$S_{11}(1535)$, respectively. We will neglect higher partial waves 
baryon resonances such as $D_{13}(1520)$ since we are only interested 
in rough qualitative predictions.
In order to achieve better agreement with experiment, one has to
consider further resonances, e.g. $D_{13}(1520)$ and $S_{11}(1650)$, and
include chiral loop corrections. But this is beyond the scope of the present
investigation and the calculations are performed at tree level.

Let us first consider the spin-$1/2^+$ octet which we denote by $P$. The
octet consists of $N^*(1440), \Sigma^*(1660), \Lambda^*(1600), \Xi^*(?)$
and the effective Lagrangian of the $P$-wave octet coupled to the ground state
baryon octet takes the form
\beq
{\cal L} = {\cal L}_{P} + {\cal L}_{\phi B P}
\eeq
with the kinetic term
\beq
{\cal L}_{P} = 
i \langle \bar{P}  \gamma_{\mu}  [D^{\mu},P] \rangle 
 - M_P \langle \bar{P}P \rangle .
\eeq
Since for the processes considered here only $N^*(1440)$ contributes, we set
$M_P = 1.44$ GeV.
The interaction terms of the $P$-wave resonances with the ground state baryon
octet read
\beqa  \label{res1}
{\cal L}_{\phi B P} &=& - \frac{1}{2} D_P \langle \bar{P} \gamma_{\mu}
 \gamma_5 \{u^{\mu},B\} \rangle  
 - \frac{1}{2} F_P \langle \bar{P} \gamma_{\mu} \gamma_5 [u^{\mu},B] 
 \rangle - \lambda_P \langle \bar{P} \gamma_{\mu} \gamma_5 B \rangle 
  \langle u^{\mu} \rangle \no \\
&&
+ d_P \langle \bar{P} \sigma^{\mu \nu} \{ F^+_{\mu \nu}, B \} \rangle 
+ f_P \langle \bar{P} \sigma^{\mu \nu} [ F^+_{\mu \nu}, B ] \rangle 
+ \mbox{h.c.} \; .
\eeqa
A possible $\eta_0 \bar{P} \gamma_5 B$ 
term can again be eliminated by using the
equation of motion for baryons.
The coupling constants $D_P, F_P$ and $d_P, f_P$ can be determined from 
strong and radiative decays of the $N^*(1440)$ resonance, \cite{B2},
\beqa  \label{rres}
D_P &=& 0.32 \pm 0.05,  \qq  F_P = 0.16  \pm 0.04 \no \\
d_P &=& -0.05 \pm 0.02 \; \mbox{GeV}^{-1}  ,  \qq  
f_P = 0.08 \pm 0.02 \; \mbox{GeV}^{-1}   ,
\eeqa
where the first number indicates the central value used in Ref. \cite{B2}
and the error bars are due to the uncertainty in the decay widths.
In $\eta$ and $\eta'$ photoproduction it has been found that the effect of
$\lambda_P$ is negligible, therefore we set it zero in our calculations: 
$\lambda_P =0$.
The spin-$1/2^-$ octet consists of $N^*(1535), \Lambda^*(1670), \Sigma^*(1750),
\Xi^*(?)$ and the pertinent Lagrangian reads
\beq
{\cal L} = {\cal L}_{S} + {\cal L}_{\phi B S}
\eeq
with the kinetic term
\beq
{\cal L}_{S} = 
i \langle \bar{S}  \gamma_{\mu}  [D^{\mu},S] \rangle 
 - M_S \langle \bar{S}S \rangle 
\eeq
and the interaction part
\beqa  \label{res2}
{\cal L}_{\phi B S} &=& - \frac{i}{2} D_S \langle \bar{S} \gamma_{\mu}
 \{u^{\mu},B\} \rangle  
 - \frac{i}{2} F_S \langle \bar{S} \gamma_{\mu} [u^{\mu},B] 
 \rangle - i \lambda_S \langle \bar{S} \gamma_{\mu} B \rangle 
  \langle u^{\mu} \rangle \no \\
&&
+ i d_S \langle \bar{S} \sigma^{\mu \nu} \gamma_5 
  \{ F^+_{\mu \nu}, B \} \rangle 
+ i f_S \langle \bar{S} \sigma^{\mu \nu} \gamma_5 
   [ F^+_{\mu \nu}, B ] \rangle 
+ \mbox{h.c.} \; .
\eeqa
We set $M_S = 1.535$ GeV and from strong and radiative decays of the $S$-wave
resonances one obtains, \cite{B2},
\beqa  \label{sres}
D_S &=& 0.37 \pm 0.06,  \qq  F_S = -0.21 \pm 0.04,  \qq 
\lambda_S = -0.07  \pm 0.02 \no \\
d_S &=& -0.07 \pm 0.03\; \mbox{GeV}^{-1}  ,  \qq  
f_S = -0.06 \pm 0.03\; \mbox{GeV}^{-1}   .
\eeqa
Since there exists data on decay channels of the $S$-wave resonances 
into $\eta$,
we are able to fix the coupling $\lambda_S$ by taking $\eta$-$\eta'$ mixing
into account.
Two remarks are in order.
First, we would like to point out that our simple ansatz of zero width
resonances will lead to singularities at the resonance mass which
could be circumvented by the use of a finite width.
This will restrict in the case of $\eta$ electroproduction
the validity of our approach to energies very close to threshold
which we are considering in the present work,
whereas it is numerically irrelevant for $\eta'$ electroproduction.
Second, we have calculated both Born terms using the lowest order chiral
effective Lagrangian and resonance contributions. We would like to emphasize
that this procedure does not imply any double counting. The contributions of
the resonances are hidden only in higher chiral order \mbox{counterterms}
of the effective Lagrangian which we did not take into account in the present
investigation. Born terms like the ones used in this work are not produced by
resonance contributions.

\section{Invariant amplitudes and numerical results}
We proceed by presenting  the invariant amplitudes for $\eta$ and $\eta'$
photoproduction on the nucleons.
Let us start with the Born terms which are depicted in
Fig. 1. They read for the proton
\beqa
A_1(p \gamma^* \rightarrow p \phi)  &=& - 2 M_N e A_\phi
   \Big[ \frac{1}{s - M_N^2} + \frac{1}{u - M_N^2} + \frac{\mu_p}{2 M_N^2} 
    \Big] \no \\
A_2(p \gamma^* \rightarrow p \phi)  &=&  4 M_N e A_\phi
   \frac{s+u - 2 M_N^2}{[s - M_N^2][u - M_N^2][m_\phi^2 -t]} \no \\
A_3(p \gamma^* \rightarrow p \phi)  &=& 
  e \mu_p A_\phi \Big[ \frac{1}{s - M_N^2} - \frac{1}{u - M_N^2} \Big] \no \\
A_4(p \gamma^* \rightarrow p \phi)  &=& 
  e \mu_p A_\phi \Big[ \frac{1}{s - M_N^2} + \frac{1}{u - M_N^2} \Big] \no \\
A_5(p \gamma^* \rightarrow p \phi)  &=&  -2 M_N e A_\phi
   \frac{s-u}{[s - M_N^2][u - M_N^2][m_\phi^2 -t]} \no \\
A_6(p \gamma^* \rightarrow p \phi) &=& 0 
\eeqa
with $\mu_p = 1 + b_6^F + b_6^D/3 =2.57$ being the magnetic moment of the
proton at lowest chiral order and
\beqa
A_\eta &=& \frac{1}{2 \sqrt{3} F_\pi} [D-3F] \cos \theta + \sqrt{\frac{2}{3}}
  \, \frac{1}{F_0}  [D+3 \lambda] \sin \theta   \no \\
A_{\eta'} &=&\frac{1}{2 \sqrt{3} F_\pi} [D-3F] \sin \theta - \sqrt{\frac{2}{3}}
  \, \frac{1}{F_0}  [D+3 \lambda] \cos \theta   .
\eeqa
In the case of the neutron the photon couples only via the magnetic moment and
the pertinent contribution reads
\beqa
A_1(n \gamma^* \rightarrow n \phi)  &=& 
  - \frac{e}{ M_N} \mu_n A_\phi \no \\
A_3(n \gamma^* \rightarrow n \phi)  &=& 
  e \mu_n A_\phi \Big[ \frac{1}{s - M_N^2} - \frac{1}{u - M_N^2} \Big] \no \\
A_4(n \gamma^* \rightarrow n \phi)  &=& 
  e \mu_n A_\phi \Big[ \frac{1}{s - M_N^2} + \frac{1}{u - M_N^2} \Big] \no \\
A_2(n \gamma^* \rightarrow n \phi) &=&  
A_5(n \gamma^* \rightarrow n \phi) \q = \q
A_6(n \gamma^* \rightarrow n \phi) = 0 
\eeqa
with $\mu_n = - 2 b_6^D/3 = -1.59$ being the neutron magnetic moment 
at lowest chiral order.

Vector meson exchange is shown in Fig. 2. One has to add the
following terms to the invariant amplitudes for photoproduction on the proton
\beqa
A_1(p \gamma^* \rightarrow p \phi)  &=& \frac{e \kappa_\rho}{4 M_N} g_{\rho N} 
  \, g_{\rho \gamma \phi} \frac{t}{t - M_\rho^2}  \no \\
A_2(p \gamma^* \rightarrow p \phi)  &=& \frac{e \kappa_\rho}{4 M_N} g_{\rho N} 
  \, g_{\rho \gamma \phi} \frac{t - m_\phi^2 + k^2}{[t - M_\rho^2]
      [t -m_\phi^2] }  \no \\
A_4(p \gamma^* \rightarrow p \phi)  &=&  - \frac{e}{2}  
 \sum_{V= \rho_0, \omega} g_{V N}\,  
  g_{V \gamma \phi} \frac{1}{t - M_V^2} \no \\
A_5(p \gamma^* \rightarrow p \phi)  &=& - \frac{e \kappa_\rho}{8 M_N} 
  g_{\rho N} \, g_{\rho \gamma \phi} \frac{s-u}{[t - M_\rho^2]
      [t -m_\phi^2] }  \no \\
A_3(p \gamma^* \rightarrow p \phi)  &=&  
A_6(p \gamma^* \rightarrow p \phi) \q = \q 0. 
\eeqa
For the neutron $g_{\rho N}$ has to be replaced by $-g_{\rho N}$.

We now turn to the baryon resonances. Their contributions are given in Fig. 3
and read for the spin-1/2$^+$ octet in the proton case
\beqa \label{probar} 
A_1(p \gamma^* \rightarrow p \phi)  &=&  - e \frac{4}{3} (d_P + 3 f_P) P_\phi
\Big[ \frac{u - M_N^2}{u- M_P^2} + \frac{s - M_N^2}{s- M_P^2} \Big] \no \\
A_3(p \gamma^* \rightarrow p \phi)  &=&   e \frac{4}{3} (d_P + 3 f_P) P_\phi
( M_P + M_N) \Big[ \frac{1}{s- M_P^2} - \frac{1}{u- M_P^2} \Big] \no \\
A_4(p \gamma^* \rightarrow p \phi)  &=&   e \frac{4}{3} (d_P + 3 f_P) P_\phi
( M_P + M_N) \Big[ \frac{1}{s- M_P^2} + \frac{1}{u- M_P^2} \Big] \no \\
A_2(p \gamma^* \rightarrow p \phi)  &=& 
A_5(p \gamma^* \rightarrow p \phi) \q =\q
A_6(p \gamma^* \rightarrow p \phi) \q =\q   0 
\eeqa
with
\beqa
P_\eta &=& \frac{1}{2 \sqrt{3} F_\pi} [D_P-3F_P] \cos \theta + 
\sqrt{\frac{2}{3}}  \, \frac{1}{F_0}  [D_P+3 \lambda_P] \sin \theta   \no \\
P_{\eta'} &=& \frac{1}{2 \sqrt{3} F_\pi} [D_P-3F_P] \sin \theta - 
\sqrt{\frac{2}{3}}  \, \frac{1}{F_0}  [D_P+3 \lambda_P] \cos \theta , 
\eeqa
whereas the results for the neutron are obtained by replacing $d_P + 3 f_P$ by
$-2d_P$ in Eq. (\ref{probar}).
The contributions from the spin-1/2$^-$ resonances read in the case of the
proton 
\beqa \label{barpro}
A_1(p \gamma^* \rightarrow p \phi)  &=&   e \frac{4}{3} (d_S + 3 f_S) S_\phi
\Big[ \frac{u - M_N^2}{u- M_S^2} + \frac{s - M_N^2}{s- M_S^2} \Big] \no \\
A_3(p \gamma^* \rightarrow p \phi)  &=&   e \frac{4}{3} (d_S + 3 f_S) S_\phi
( M_S - M_N) \Big[ \frac{1}{s- M_S^2} - \frac{1}{u- M_S^2} \Big] \no \\
A_4(p \gamma^* \rightarrow p \phi)  &=&   e \frac{4}{3} (d_S + 3 f_S) S_\phi
( M_S - M_N) \Big[ \frac{1}{s- M_S^2} + \frac{1}{u- M_S^2} \Big] \no \\
A_2(p \gamma^* \rightarrow p \phi)  &=&  
A_5(p \gamma^* \rightarrow p \phi)  \q = \q
A_6(p \gamma^* \rightarrow p \phi)  \q = \q 0  
\eeqa
with
\beqa
S_\eta &=& \frac{1}{2 \sqrt{3} F_\pi} [D_S-3F_S] \cos \theta + 
\sqrt{\frac{2}{3}}  \frac{1}{F_0}  [D_S+3 \lambda_S] \sin \theta   \no \\
S_{\eta'} &=& \frac{1}{2 \sqrt{3} F_\pi} [D_S-3F_S] \sin \theta - 
\sqrt{\frac{2}{3}}  \frac{1}{F_0}  [D_S+3 \lambda_S] \cos \theta ,
\eeqa
where for neutrons $d_S + 3 f_S $ in Eq. (\ref{barpro})
has to be replaced by $-2d_S$.

\subsection{Numerical results}
In this subsection, we discuss the numerical results for the 
separated differential cross sections.
From semileptonic decays of the ground state
baryon octet and from strong and radiative decays of the baryon resonances
one can determine most LECs. The
coupling constants of the vector meson Lagrangian are quite well known. 
For the remaining parameters we take the values from the discussion in
Sec. 3 and for $F_0$ we employ the large $N_c$ identity $F_0 =
F_\pi$.
In Figures 4 to 7 the separated differential cross sections are given
at $k^2 = -0.04$ GeV$^2$, $s= 2.215$ GeV$^2$ and $s= 3.604$ GeV$^2$ for 
$\eta$ and $\eta'$ electroproduction, respectively,
where we have chosen the central values of the parameters.
The dominance of the $S$-wave multipoles can clearly be seen from the
differential cross sections $d \sigma_T/d \Omega$ and $d \sigma_L/d \Omega$ 
which remain almost constant for different angles $\vartheta$.
Although no experimental data exists at present for $\eta$ and $\eta'$
electroproduction close to threshold and small momentum transfer we do not
present results for higher values of $s$ and $|k^2|$ 
since further resonances and
loop contributions will start dominating.
Therefore, it remains to be seen if our simple model is capable of reproducing
experimental data close to threshold and small momentum transfer.

It is also worth estimating the theoretical uncertainty within our tree level
model. Note that we do not consider the errors which arise from neglecting
further resonances and loop diagrams.
The only uncertainty stems then from a variation of the resonance couplings. We
restrict ourselves to a discussion of the separated differential cross sections
at $\cos \theta =0$ since the dependence on the different resonance couplings
at one scattering angle is indicative for all other angles.
We vary both the vector resonance couplings
$g_{V \gamma \eta}$, $g_{V \gamma \eta'}$ from Eqs. (\ref{gvge}) and 
(\ref{gvgep}), and the baryon resonance couplings in  Eqs. (\ref{rres}) and
(\ref{sres}). It turns out that for a variation of the couplings of both the
vector mesons and $P$-wave baryon resonances within their ranges as given in
Sec. 3 the separated differential cross sections change less than $15 \%$.
E.g., varying $g_{\rho \gamma \eta}$ from $1.30$ GeV$^{-1}$ up to
$1.62$ GeV$^{-1}$ while keeping the remaining couplings fixed leads 
(at $\cos \theta =0$) to 
$d \sigma_T / d \Omega (p \gamma^* \rightarrow p \eta)=
0.53 \mu b/sr$ and $0.61\mu b/sr$, respectively, and for the central value of
 $g_{\rho \gamma \eta} =1.46$ GeV$^{-1}$ one obtains 
$d \sigma_T / d \Omega (p \gamma^* \rightarrow p \eta) = 0.57 \mu b/sr$.
Even smaller changes occur when the parameters for the $P$-wave resonances
are varied within their error bars. Choosing, e.g., $D_P =0.37$,
$F_P =0.12$ and the central values for the other parameters
leads to $d \sigma_T / d \Omega (p \gamma^* \rightarrow p 
\eta) = 0.56 \mu b/sr$.
One observes a similar behavior also for the other separated differential cross
sections and in the case of the neutron.
A variation of the $S$-wave resonance couplings, on the other hand, leads to
substantial changes in the numerical results. This is again a confirmation of
the $S$-wave multipole dominance.
In order to estimate the range the separated differential cross
sections can occupy when varying the resonance couplings, it is therefore
sufficient to change the $S$-wave resonance couplings
$D_S, F_S, \lambda_S, d_S$ and $f_S$ while keeping the remaining couplings
fixed at their central values.
The two sets of parameters 
$D_S=0.31 , F_S=-0.17, \lambda_S=-0.05, d_S = -0.04$ GeV$^{-1}$, 
$f_S = -0.04$ GeV$^{-1}$
and $D_S=0.43 , F_S=-0.25, \lambda_S=-0.09, d_S = -0.10$ GeV$^{-1}$, 
$f_S = -0.09$  GeV$^{-1}$
maximize the change in the separated differential cross sections at
$\cos \theta =0$. One obtains in units of $\mu b/sr$ for the proton
\beqa
\frac{d \sigma_T}{d \Omega} (p \gamma^* \rightarrow p \eta)   &=&
0.25 \dots 1.33 , \q (0.57) \no \\
\frac{d \sigma_L}{d \Omega} (p \gamma^* \rightarrow p \eta)   &=&
0.35 \dots 1.03 , \q (0.57) \no \\
\frac{d \sigma_{TL}}{d \Omega} (p \gamma^* \rightarrow p \eta)   &=&
[0.39 \dots 0.61] \times 10^{-1} , \q (0.48 \times 10^{-1} ) \no \\
\frac{d \sigma_{TT}}{d \Omega} (p \gamma^* \rightarrow p \eta)   &=&
[0.31 \dots 0.30] \times 10^{-2} , \q (0.31 \times 10^{-2} ) \no \\
\frac{d \sigma_T}{d \Omega} (p \gamma^* \rightarrow p \eta')   &=&
0.48 \dots 0.80 , \q (0.60) \no \\
\frac{d \sigma_L}{d \Omega} (p \gamma^* \rightarrow p \eta')   &=&
0.37 \dots 0.54 , \q (0.43) \no \\
\frac{d \sigma_{TL}}{d \Omega} (p \gamma^* \rightarrow p \eta')   &=&
[0.37 \dots 0.45] \times 10^{-1} , \q (0.40 \times 10^{-1} ) \no \\
\frac{d \sigma_{TT}}{d \Omega} (p \gamma^* \rightarrow p \eta')   &=&
[0.31 \dots 0.32] \times 10^{-2} , \q (0.31 \times 10^{-2} ) 
\eeqa
and in the case of the neutron
\beqa
\frac{d \sigma_T}{d \Omega} (n \gamma^* \rightarrow n \eta)   &=&
0.07 \dots 0.30 , \q (0.15) \no \\
\frac{d \sigma_L}{d \Omega} (n \gamma^* \rightarrow n \eta)   &=&
[0.01 \dots 0.41] \times 10^{-1} , \q (0.10 \times 10^{-1}) \no \\
\frac{d \sigma_{TL}}{d \Omega} (n \gamma^* \rightarrow n \eta)   &=&
[-0.26 \dots -0.83] \times 10^{-2} , \q (-0.51 \times 10^{-2} ) \no \\
\frac{d \sigma_{TT}}{d \Omega} (n \gamma^* \rightarrow n \eta)   &=&
[0.16 \dots 0.16] \times 10^{-3} , \q (0.16 \times 10^{-3} ) \no \\
\frac{d \sigma_T}{d \Omega} (n \gamma^* \rightarrow n \eta')   &=&
0.10 \dots 0.16 , \q (0.12) \no \\
\frac{d \sigma_L}{d \Omega} (n \gamma^* \rightarrow n \eta')   &=&
[0.03 \dots 0.47] \times 10^{-2} , \q (0.16  \times 10^{-2}) \no \\
\frac{d \sigma_{TL}}{d \Omega} (n \gamma^* \rightarrow n \eta')   &=&
[-0.29 \dots -0.44] \times 10^{-2} , \q (-0.36 \times 10^{-2} ) \no \\
\frac{d \sigma_{TT}}{d \Omega} (n \gamma^* \rightarrow n \eta')   &=&
[0.12 \dots 0.11] \times 10^{-3} , \q (0.12 \times 10^{-3} ) .
\eeqa
The first (second) number is the result obtained by using the former (latter)
set of parameters, while the last number in brackets
is the result for the central values of the couplings.
The dependence of the separated differential cross sections on the
$S$-wave resonance couplings is substantial.
One can clearly see that this pole model provides a sensitive way of extracting
the resonance couplings for the $S$-wave resonances from $\eta$ and $\eta'$
electroproduction experiments close to threshold, whereas the uncertainties
caused by the other resonance couplings are suppressed.

\section{Summary}
We have studied $\eta$ and $\eta'$ electroproduction off protons and neutrons
in a recently proposed model, which has already been used to describe $\eta$
and  $\eta'$ photoproduction on nucleons \cite{B2}.
Within this model an effective chiral $U(3)$ Lagrangian is
constructed which describes the interactions of the pseudoscalar meson nonet
$(\pi,K,\eta,\eta')$ with the ground state baryon octet and low-lying
resonances. These include the vector mesons $\rho_0$ and $\omega$ in the
$t$-channel (the $\phi$ meson leads to much smaller contributions and can be
neglected for our purposes), and the $J^P = 1/2^+$ and $1/2^-$ baryon
resonances $P_{11}(1440)$ and $S_{11}(1535)$.
Most LECs of the effective Lagrangian can be determined using semileptonic
hyperon decays and both strong and radiative decays of the baryon
resonances. The couplings of the vector mesons are also quite well known.
Finally, the couplings of the axial flavor-singlet currents of both the ground
state and spin-1/2$^+$ resonance baryons have been fixed in $\eta$
and  $\eta'$ photoproduction. Our results are therefore predictions rather than
fits to experimental data.
Employing this Lagrangian we calculated Born terms 
including the magnetic moments of the nucleons and resonance exchange
diagrams. Confining ourselves to $S$- and $P$-wave multipoles 
we present the four
separated differential cross sections near threshold at photon virtuality
$k^2 = -0.04$ GeV$^2$.
Of course, we do not expect our model to be valid for higher c.m. energies and
momentum transfers away from threshold, since other effects such as
contributions from further resonances and chiral loop corrections will become
significant. 
Nevertheless, the present investigation could serve as a check for our simple
model and confirm results which have been obtained in the case of
photoproduction. It might furthermore indicate whether the $\eta'$ meson can be
included in baryon chiral perturbation theory as proposed in \cite{B}.

\section*{Acknowledgments}
The author wishes to thank N. Kaiser for useful discussions and suggestions.

\newpage

\section*{Figure captions}

\begin{enumerate}

\item[Fig.1] Shown are the Born terms for photoproduction on the proton.
             The photon is given by a wavy line. Solid and dashed lines denote
             proton and pseudoscalar mesons, respectively.  

\item[Fig.2] Vector meson exchange.  The photon is given by a wavy line.
             Solid and dashed lines denote nucleons and  pseudoscalar mesons, 
             respectively. The double line represents the vector meson. 

\item[Fig.3] Baryon resonance contributions.  
             The photon is given by a wavy line.
             Solid and dashed lines denote nucleons and  pseudoscalar mesons, 
             respectively. The double line represents the baryon resonances
             $P_{11}(1440)$ or $S_{11}(1535)$. 

\item[Fig.4] Given are the separated differential cross sections
             $d \sigma_T/ d \Omega_\eta  (a)$, 
             $d \sigma_L/ d \Omega_\eta  (b)$, 
             $d \sigma_{TL}/ d \Omega_\eta  (c)$, 
             $d \sigma_{TT}/ d \Omega_\eta  (d)$  for $\eta$
             electroproduction on the proton.

\item[Fig.5] Given are the separated differential cross sections
             $d \sigma_T/ d \Omega_{\eta'}  (a)$, 
             $d \sigma_L/ d \Omega_{\eta'} (b)$, 
             $d \sigma_{TL}/ d \Omega_{\eta'}  (c)$, 
             $d \sigma_{TT}/ d \Omega_{\eta'}  (d)$  for $\eta'$
             electroproduction on the proton.

\item[Fig.6] Given are the separated differential cross sections
             $d \sigma_T/ d \Omega_\eta  (a)$, 
             $d \sigma_L/ d \Omega_\eta (b)$, 
             $d \sigma_{TL}/ d \Omega_\eta  (c)$, 
             $d \sigma_{TT}/ d \Omega_\eta  (d)$  for $\eta$
             electroproduction on the neutron.

\item[Fig.7] Given are the separated differential cross sections
             $d \sigma_T/ d \Omega_{\eta'}  (a)$, 
             $d \sigma_L/ d \Omega_{\eta'}  (b)$, 
             $d \sigma_{TL}/ d \Omega_{\eta'}  (c)$, 
             $d \sigma_{TT}/ d \Omega_{\eta'}  (d)$  for $\eta'$
             electroproduction on the neutron.

\end{enumerate}

\newpage

\begin{center}
 
\begin{figure}[bth]
\centering
\centerline{
\epsfbox{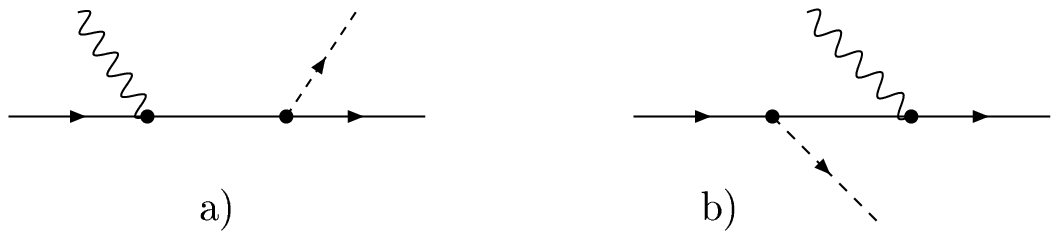}}
\end{figure}

\vskip 0.7cm

Figure 1

\vskip 2cm

\begin{figure}[tbh]
\centering
\centerline{
\epsfbox{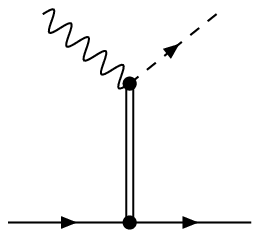}}
\end{figure}

\vskip 0.7cm

Figure 2

\vskip 2cm

\begin{figure}[tbh]
\centering
\centerline{
\epsfbox{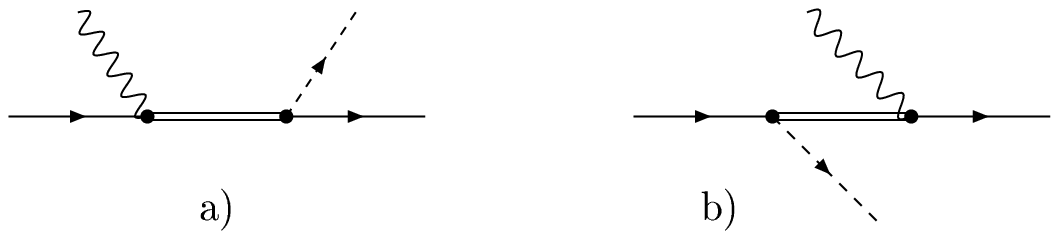}}
\end{figure}

\vskip 0.7cm

Figure 3

\begin{figure}[tbh]
\centering
\begin{picture}(300,380)  
\put(-20,260){\makebox(100,120){\epsfig{file=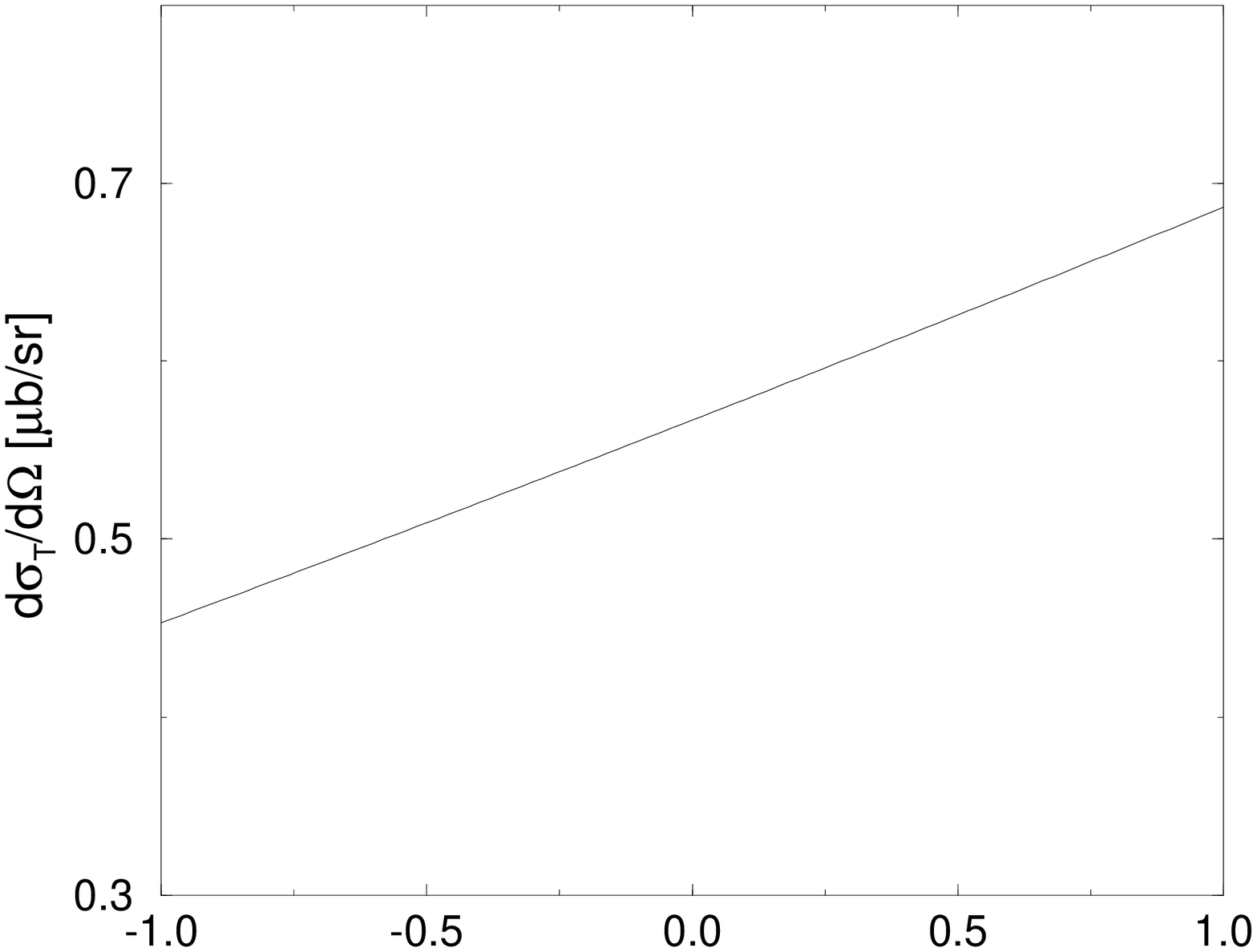,width=7.0cm,
              angle=-90}}}
\put(240,260){\makebox(100,120){\epsfig{file=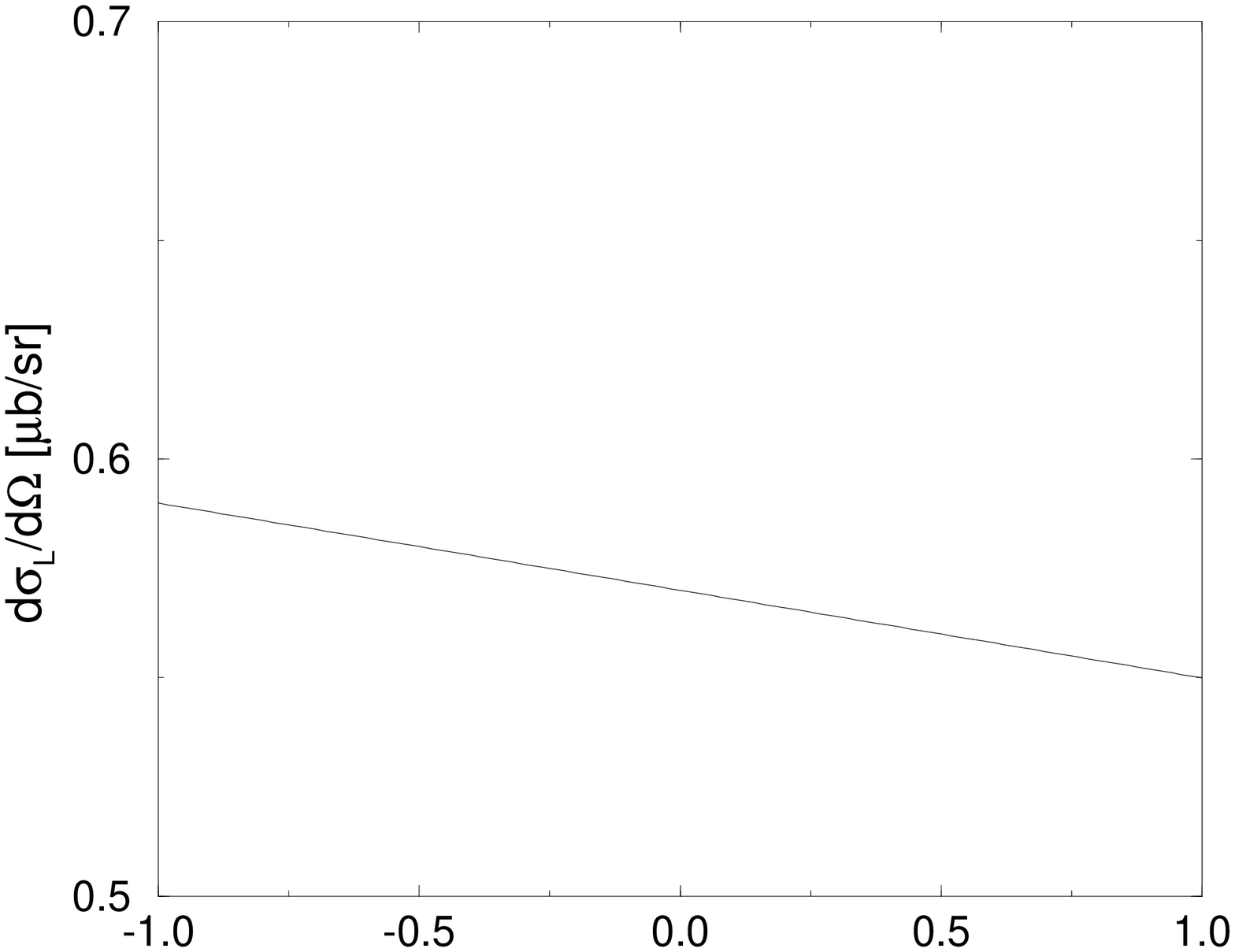,width=7.0cm,
               angle=-90}}}
\put(-20,0){\makebox(100,120){\epsfig{file=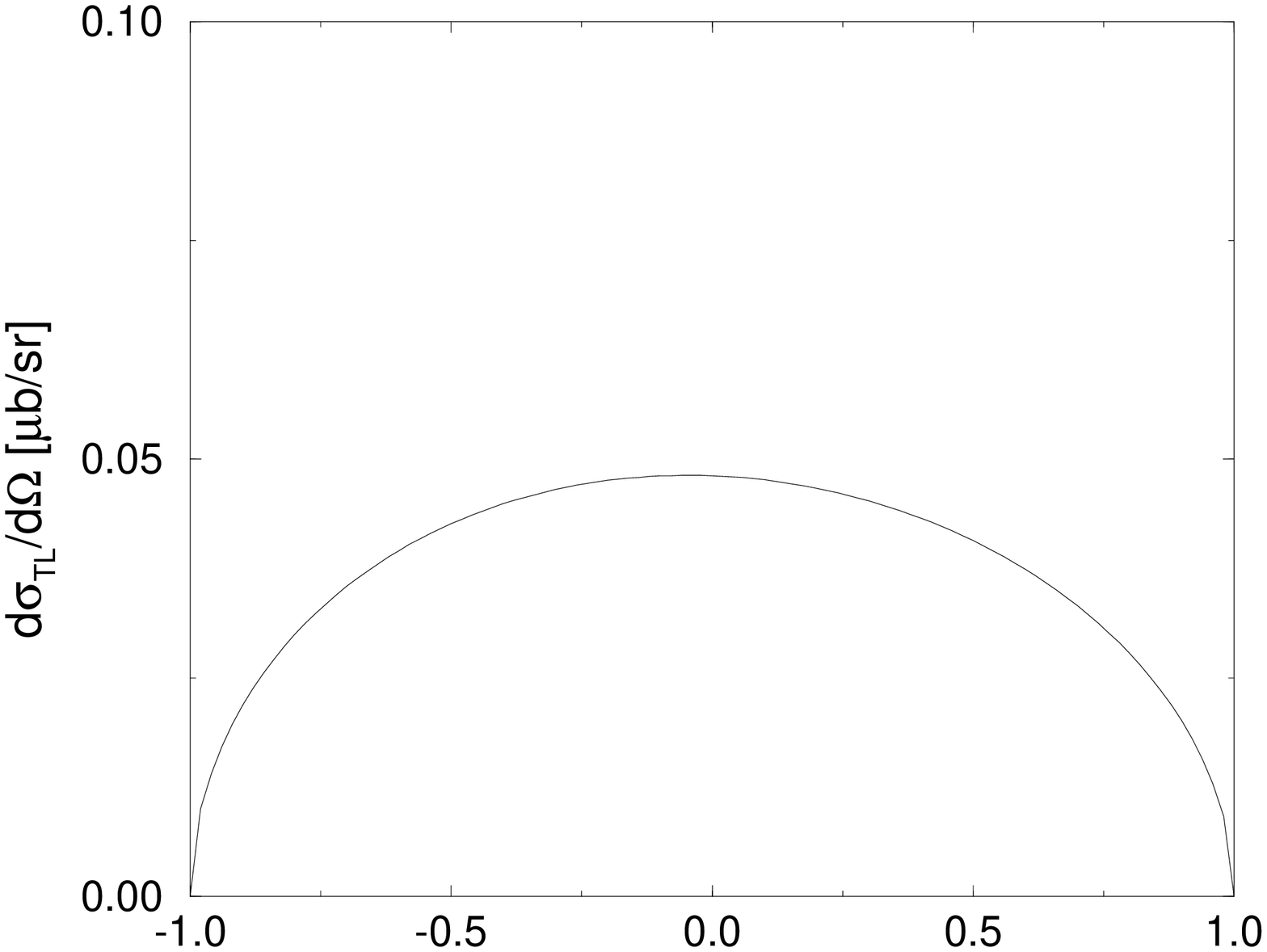,width=7.0cm,angle=-90}}}
\put(240,0){\makebox(100,120){\epsfig{file=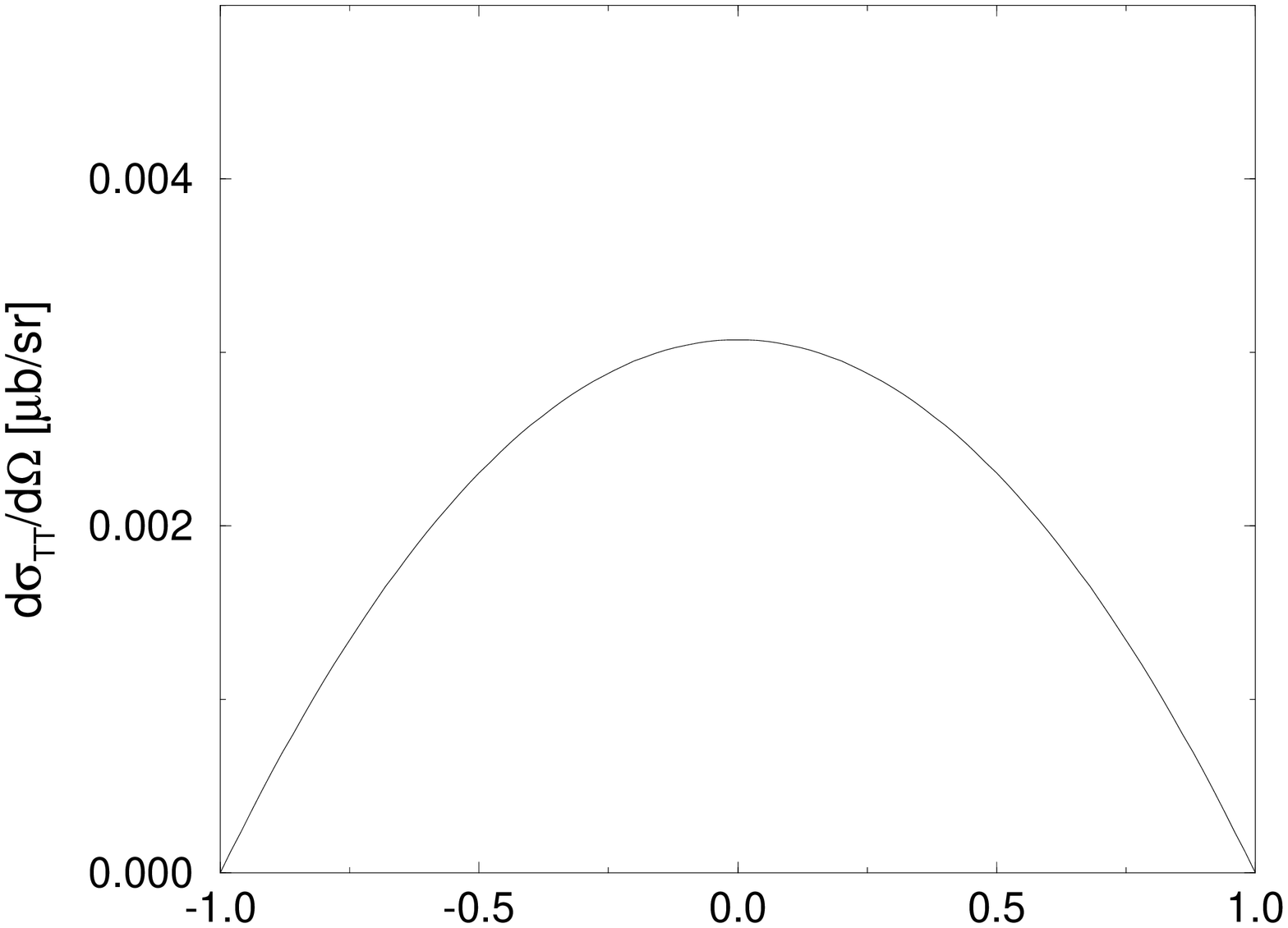,width=7.0cm,angle=-90}}}
\put(30,210){{\footnotesize $\cos \vartheta$}}
\put(35,190){$a)$}
\put(290,210){{\footnotesize $\cos \vartheta$}}
\put(295,190){$b)$}
\put(30,-50){{\footnotesize $\cos \vartheta$}}
\put(35,-70){$c)$}
\put(290,-50){{\footnotesize $\cos \vartheta$}}
\put(295,-70){$d)$}
\end{picture}
\vskip 3.8cm

Figure 4

\end{figure}

\begin{figure}[tbh]
\centering
\begin{picture}(300,380)  
\put(-20,260){\makebox(100,120){\epsfig{file=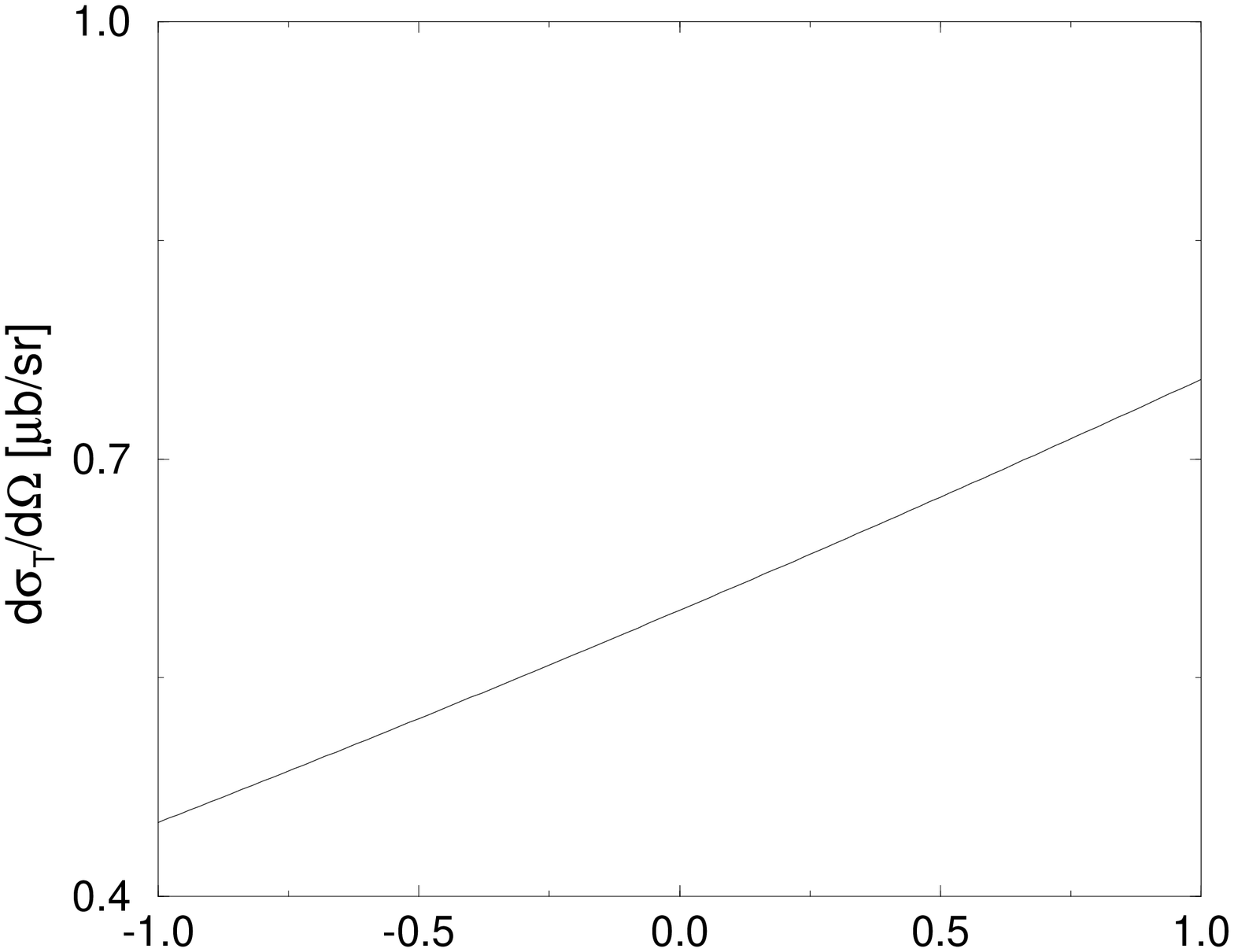,width=7.0cm,
              angle=-90}}}
\put(240,260){\makebox(100,120){\epsfig{file=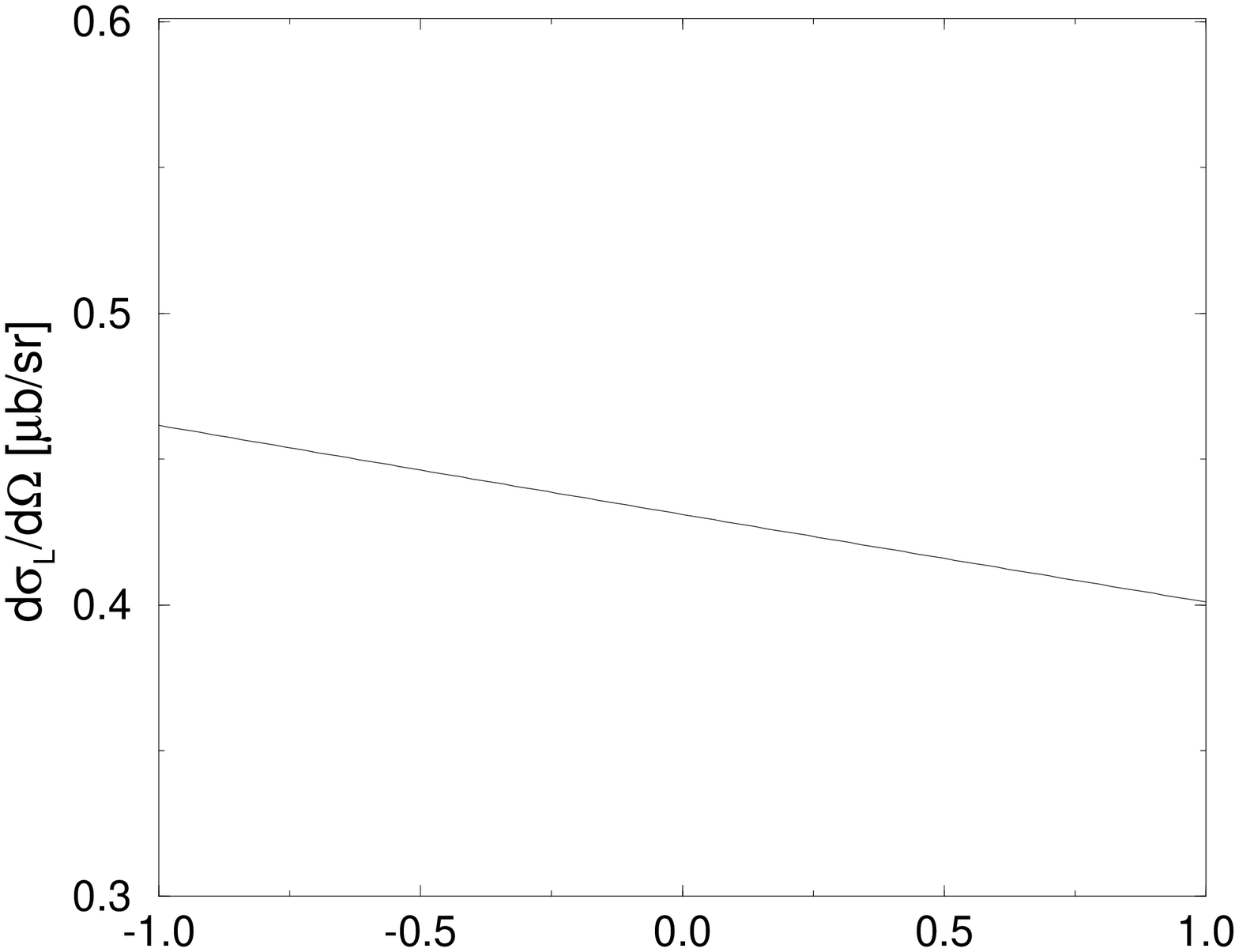,width=7.0cm,
               angle=-90}}}
\put(-20,0){\makebox(100,120){\epsfig{file=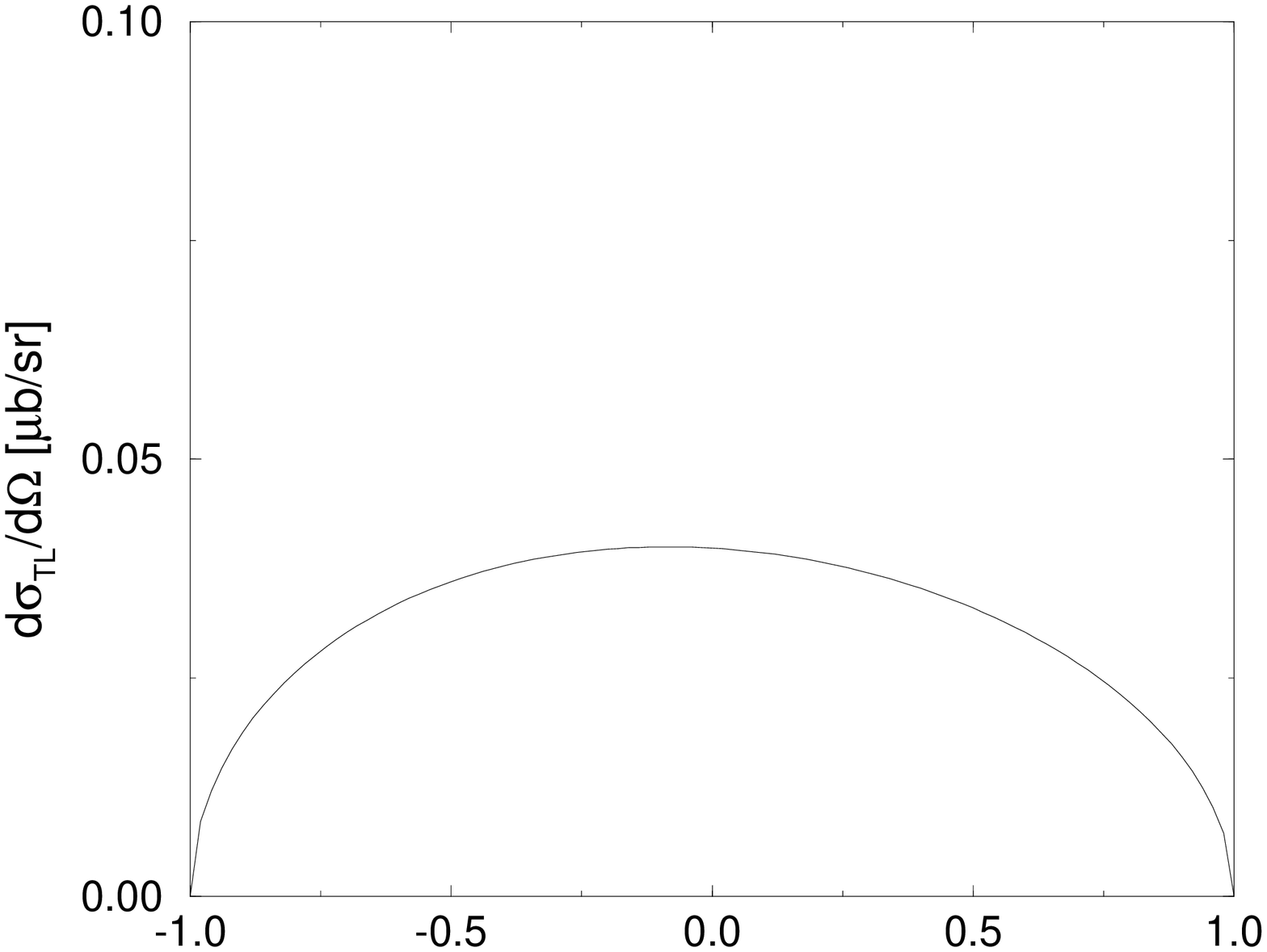,width=7.0cm,angle=-90}}}
\put(240,0){\makebox(100,120){\epsfig{file=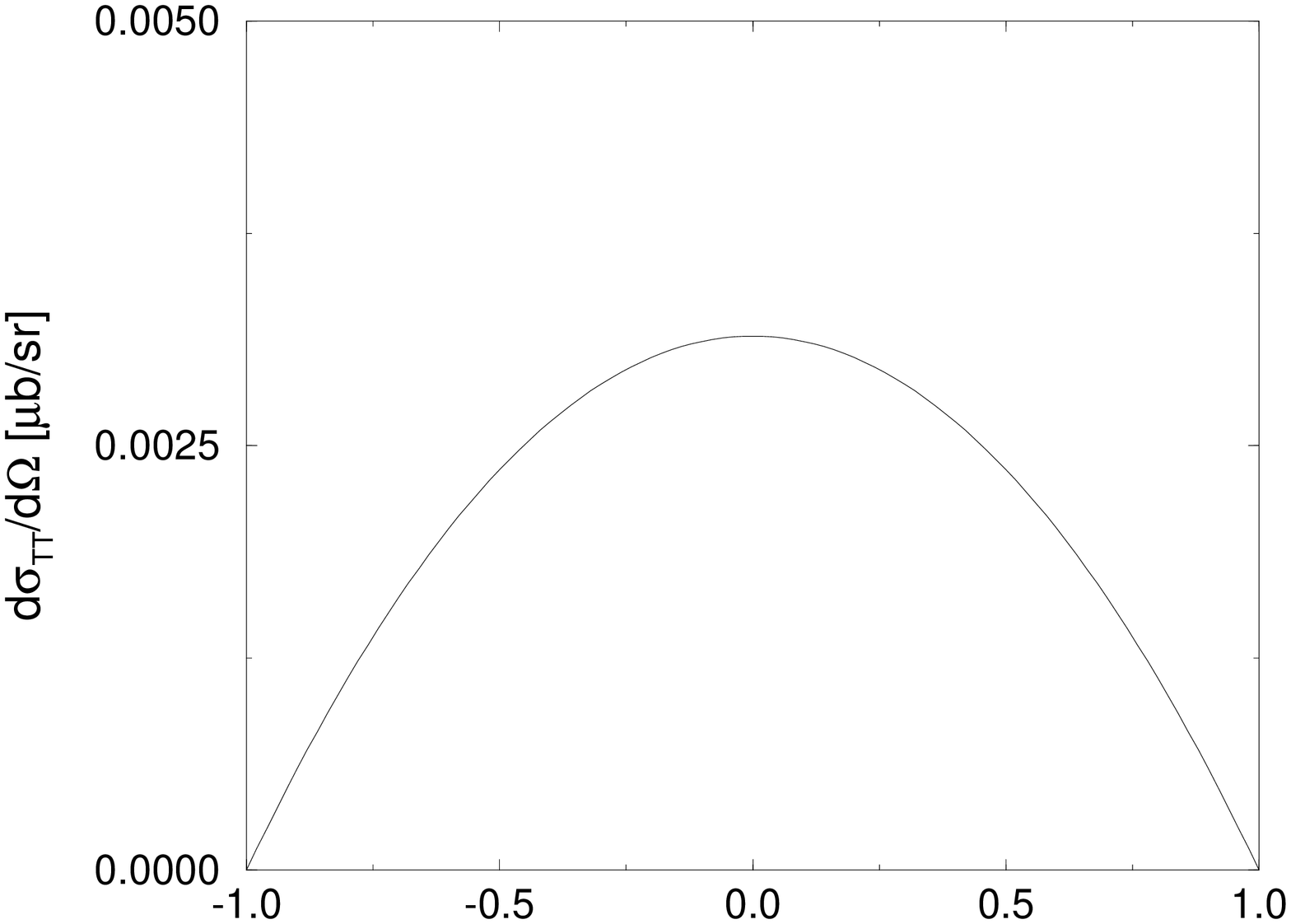,width=7.0cm,angle=-90}}}
\put(30,210){{\footnotesize $\cos \vartheta$}}
\put(35,190){$a)$}
\put(290,210){{\footnotesize $\cos \vartheta$}}
\put(295,190){$b)$}
\put(30,-50){{\footnotesize $\cos \vartheta$}}
\put(35,-70){$c)$}
\put(290,-50){{\footnotesize $\cos \vartheta$}}
\put(295,-70){$d)$}
\end{picture}
\vskip 3.8cm

Figure 5

\end{figure}

\begin{figure}[tbh]
\centering
\begin{picture}(300,380)  
\put(-20,260){\makebox(100,120){\epsfig{file=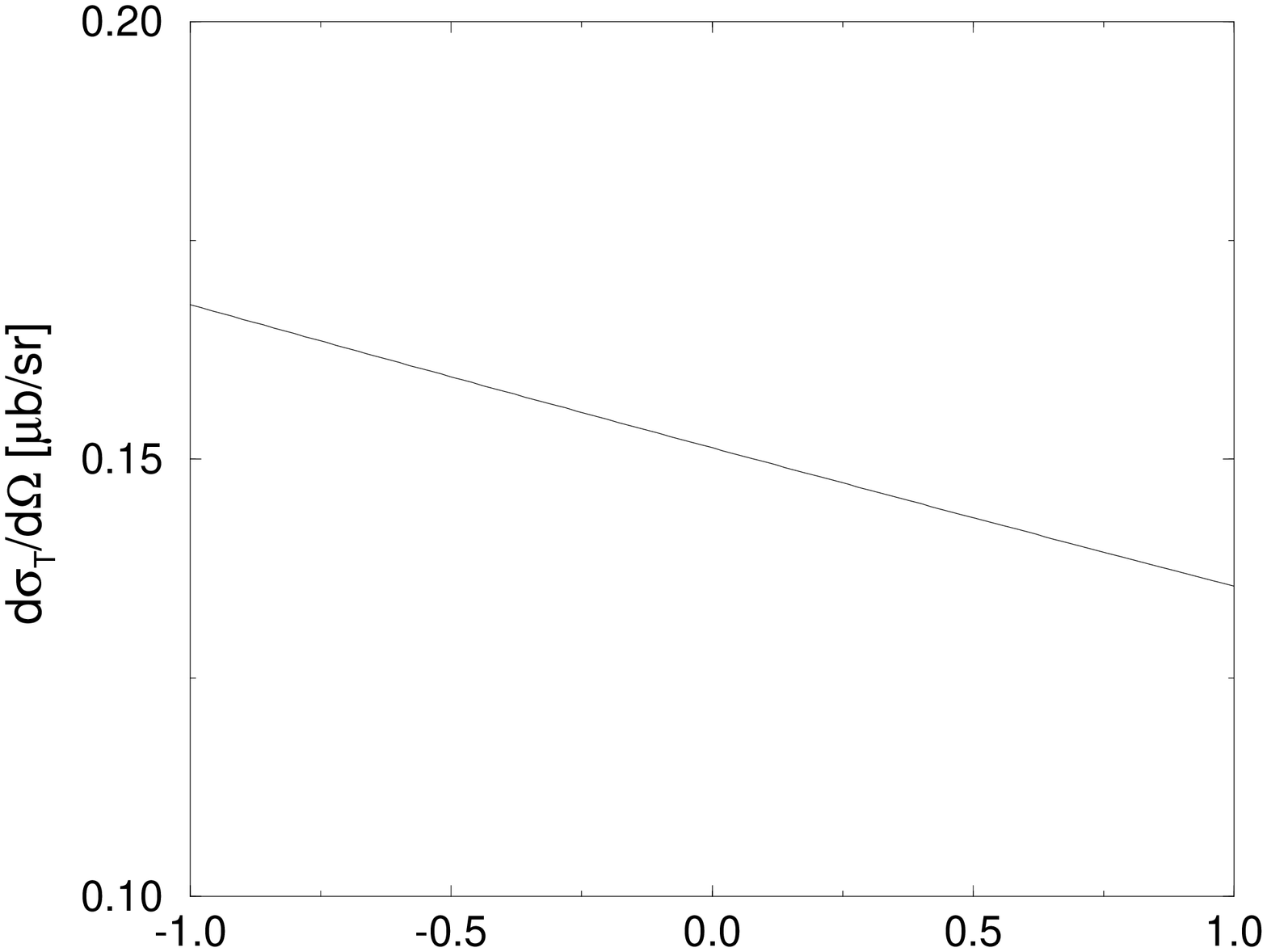,width=7.0cm,
              angle=-90}}}
\put(240,260){\makebox(100,120){\epsfig{file=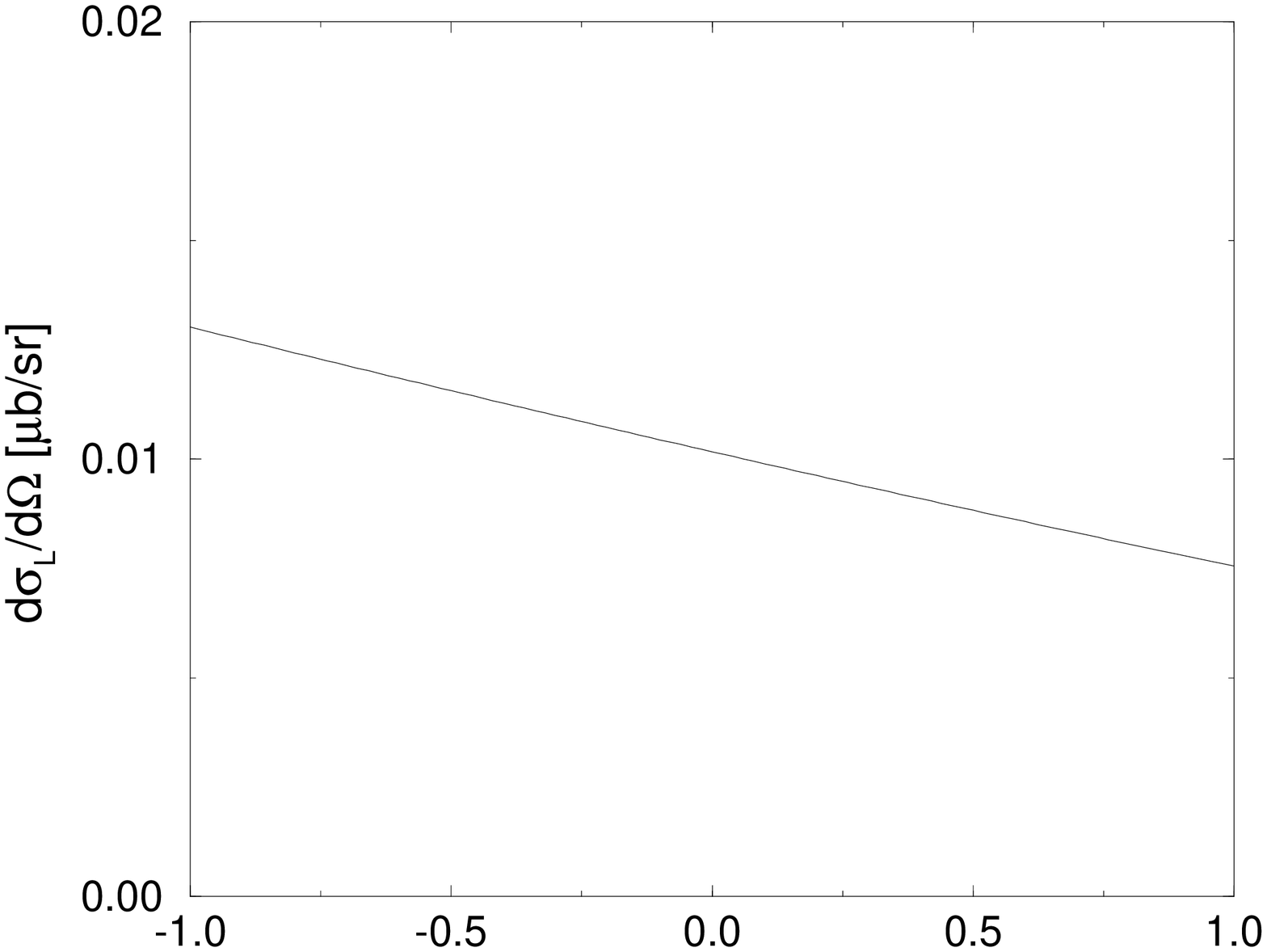,width=7.0cm,
               angle=-90}}}
\put(-20,0){\makebox(100,120){\epsfig{file=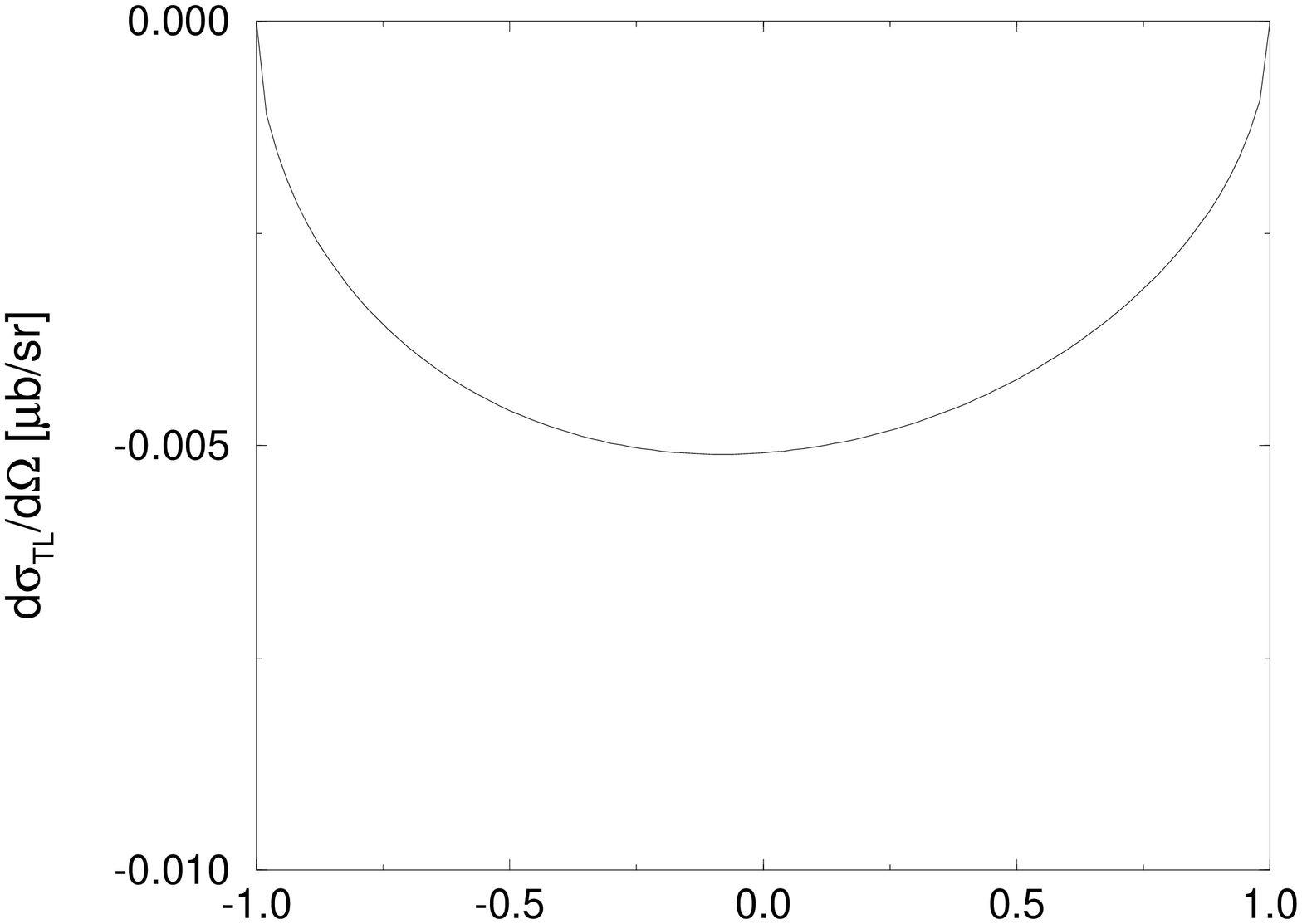,width=7.0cm,angle=-90}}}
\put(240,0){\makebox(100,120){\epsfig{file=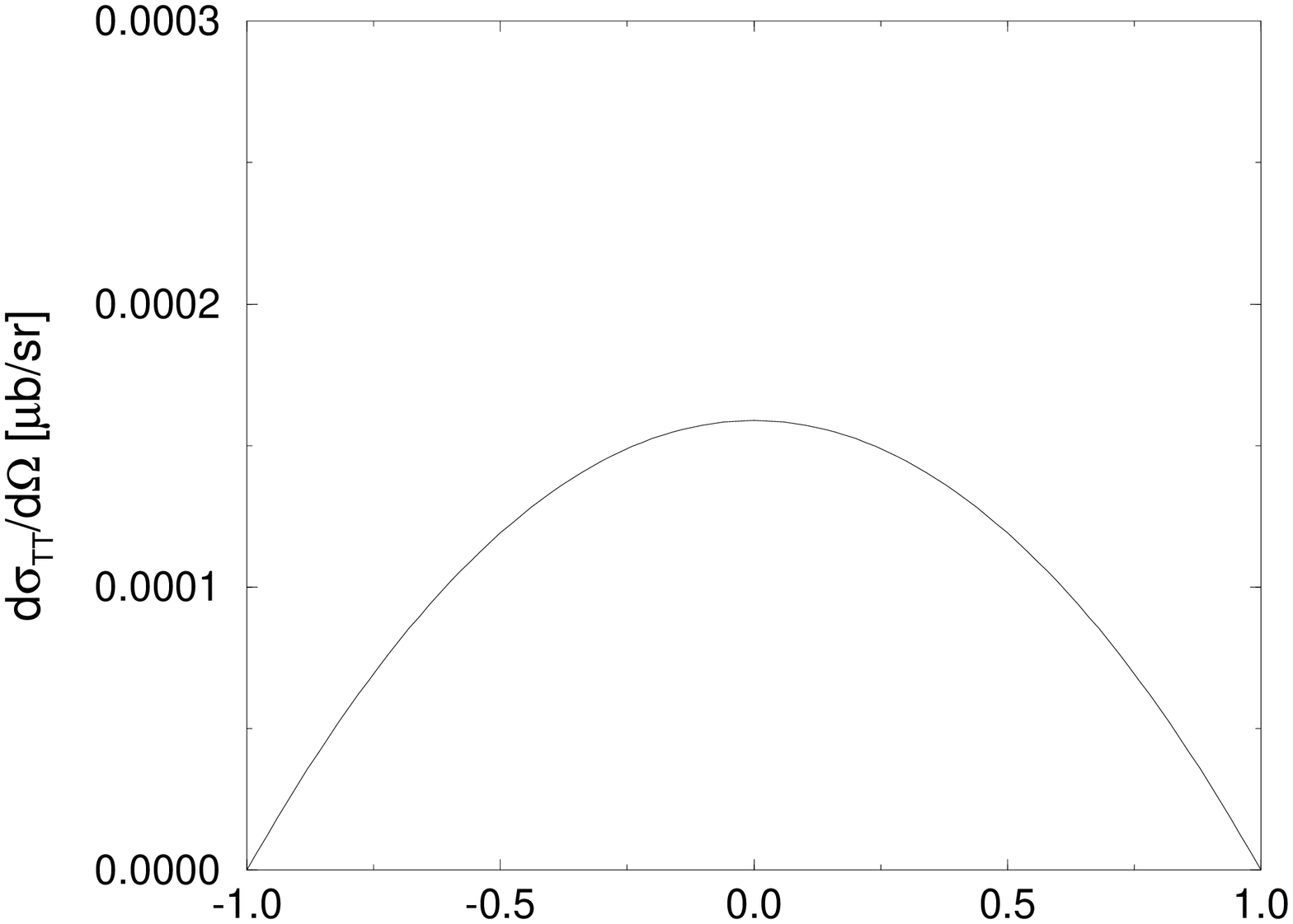,width=7.0cm,angle=-90}}}
\put(30,210){{\footnotesize $\cos \vartheta$}}
\put(35,190){$a)$}
\put(290,210){{\footnotesize $\cos \vartheta$}}
\put(295,190){$b)$}
\put(30,-50){{\footnotesize $\cos \vartheta$}}
\put(35,-70){$c)$}
\put(290,-50){{\footnotesize $\cos \vartheta$}}
\put(295,-70){$d)$}
\end{picture}
\vskip 3.8cm

Figure 6
\end{figure}

\begin{figure}[tbh]
\centering
\begin{picture}(300,380)  
\put(-20,260){\makebox(100,120){\epsfig{file=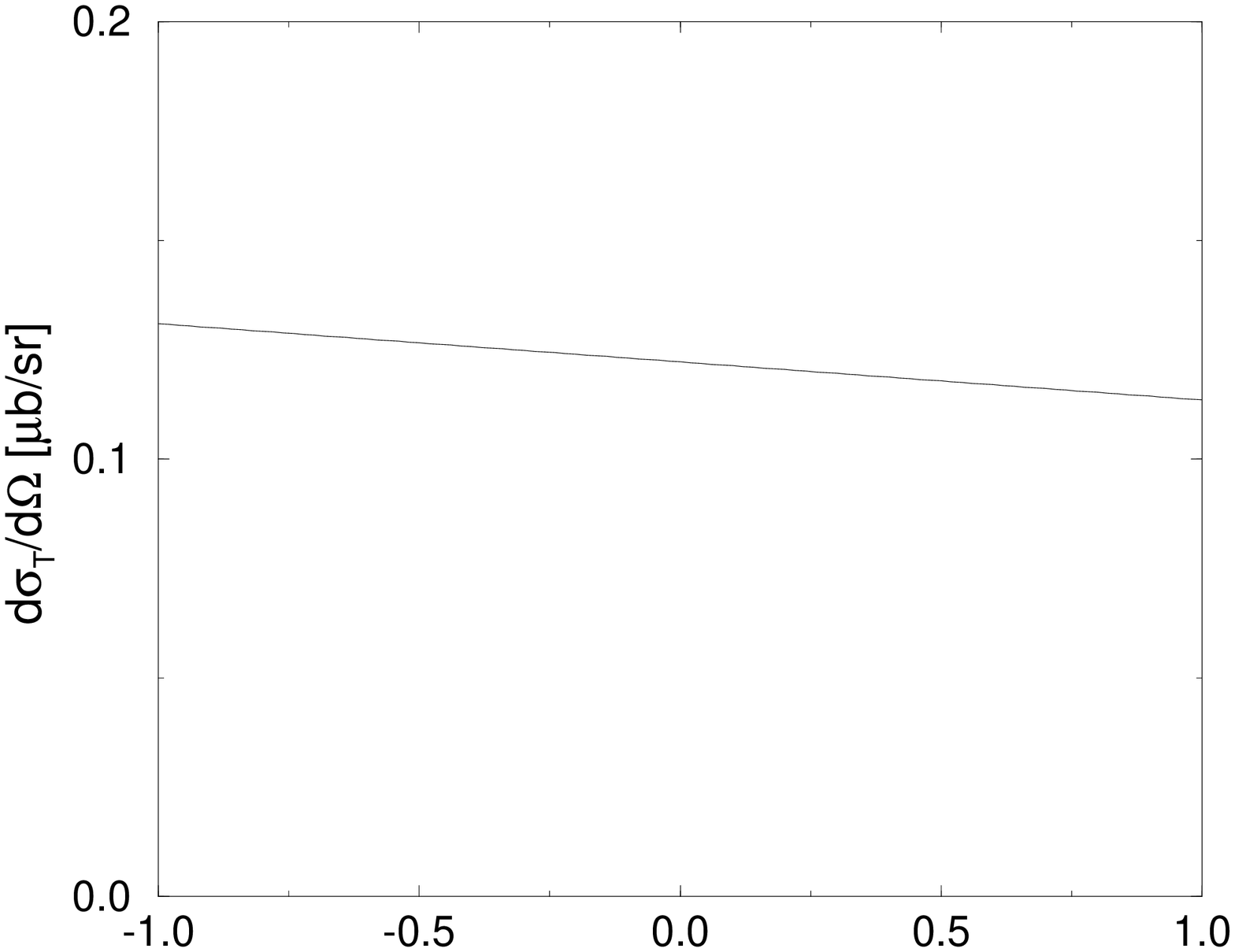,width=7.0cm,
              angle=-90}}}
\put(240,260){\makebox(100,120){\epsfig{file=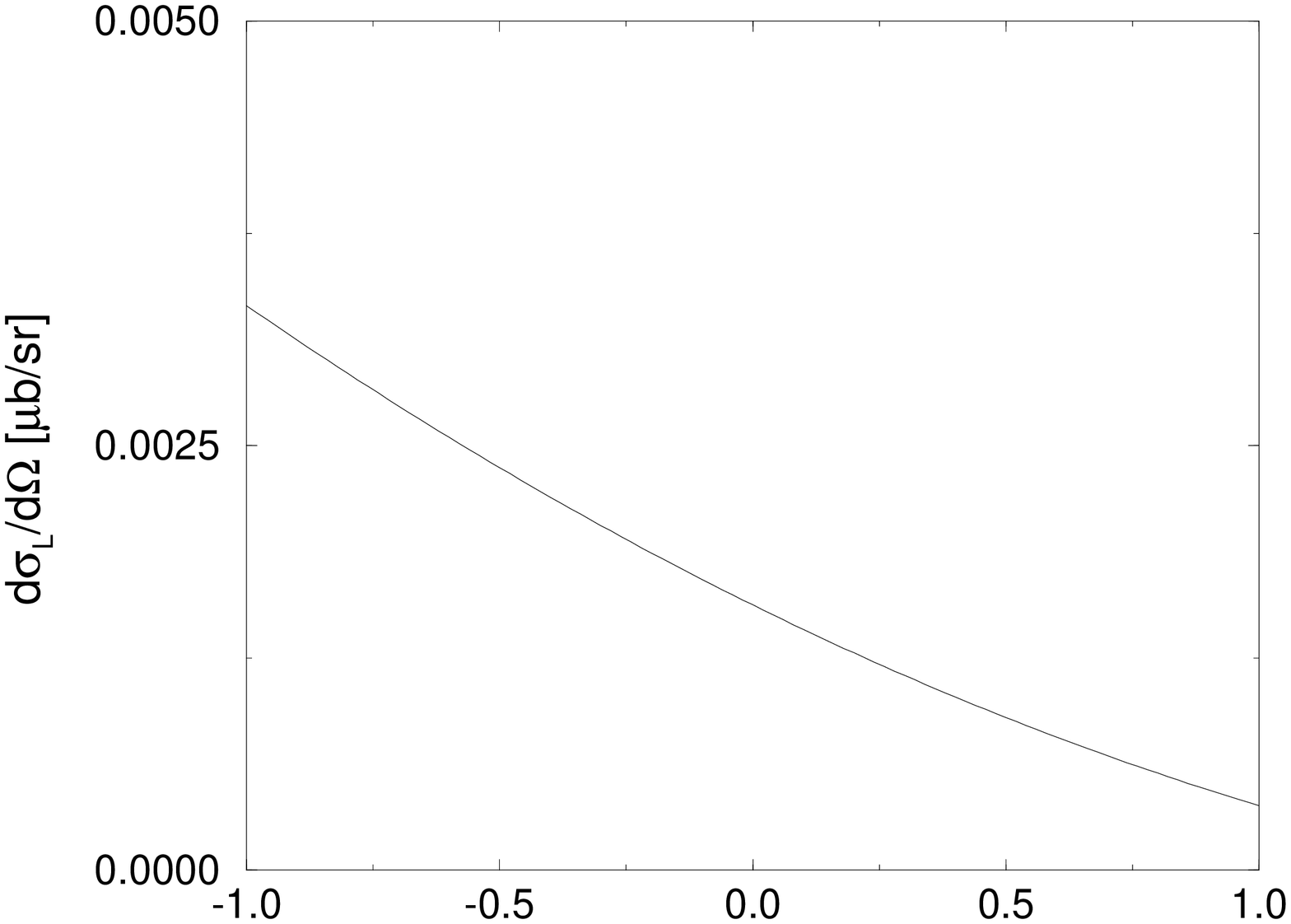,width=7.0cm,
               angle=-90}}}
\put(-20,0){\makebox(100,120){\epsfig{file=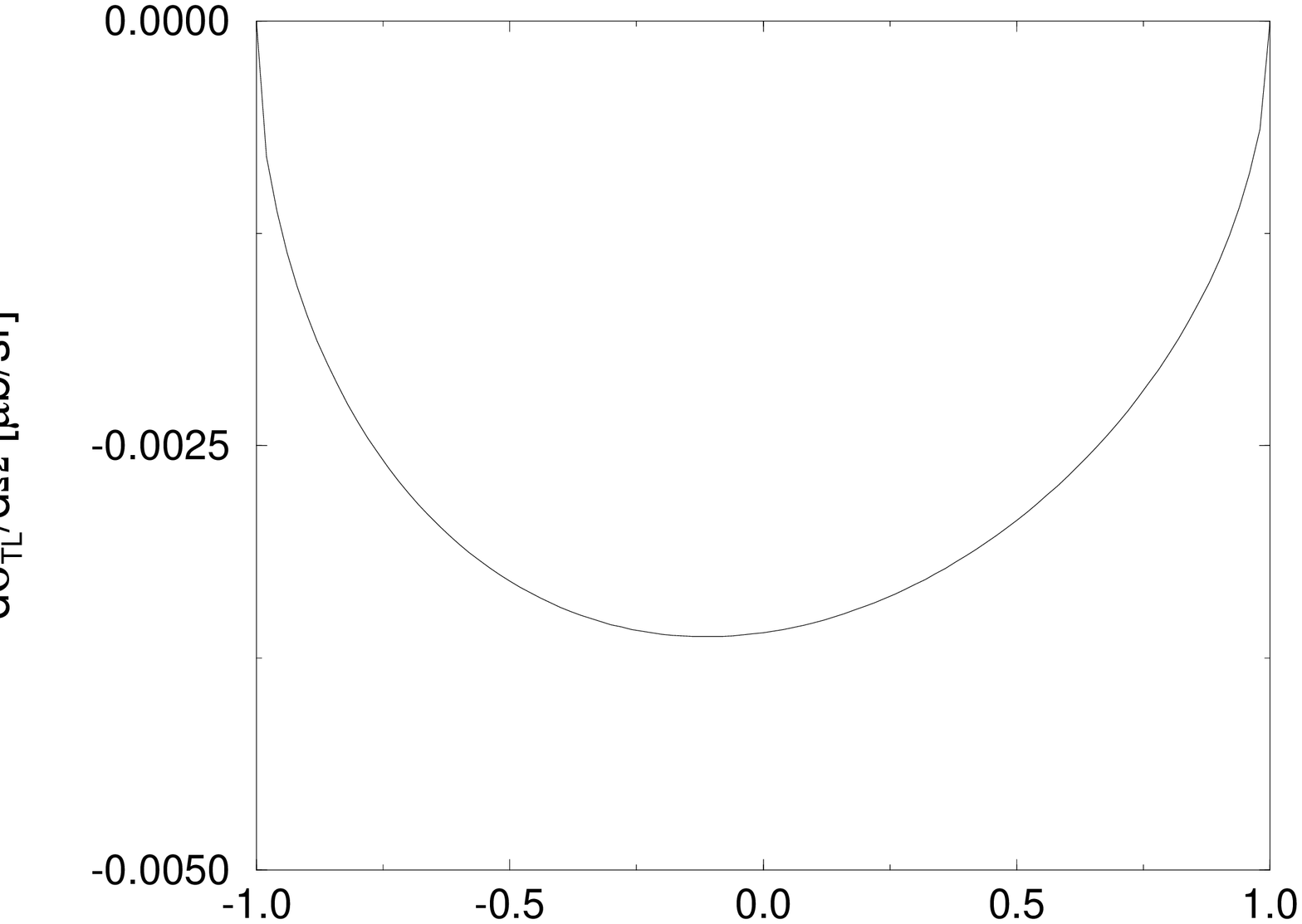,width=7.0cm,angle=-90}}}
\put(240,0){\makebox(100,120){\epsfig{file=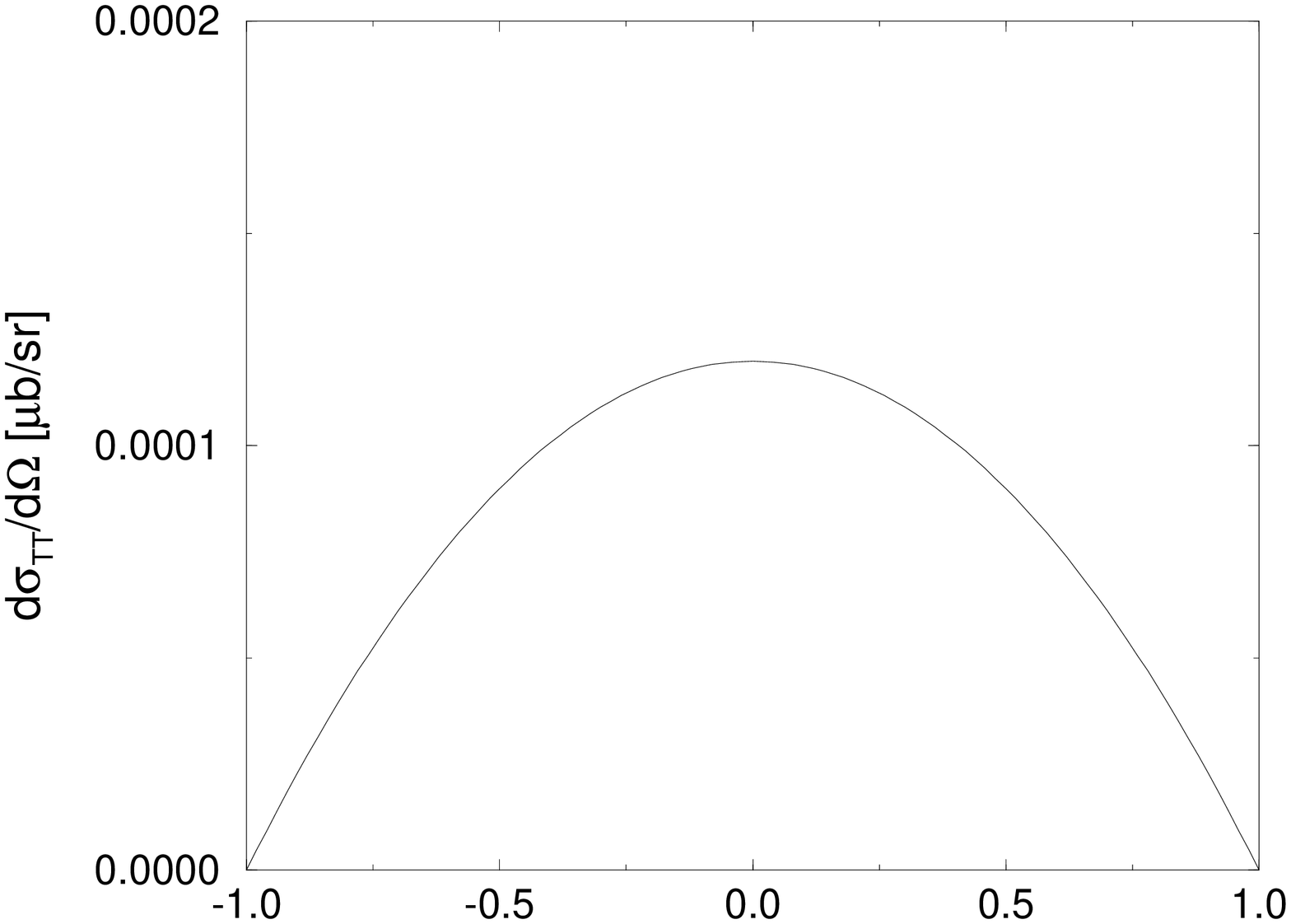,width=7.0cm,angle=-90}}}
\put(30,210){{\footnotesize $\cos \vartheta$}}
\put(35,190){$a)$}
\put(290,210){{\footnotesize $\cos \vartheta$}}
\put(295,190){$b)$}
\put(30,-50){{\footnotesize $\cos \vartheta$}}
\put(35,-70){$c)$}
\put(290,-50){{\footnotesize $\cos \vartheta$}}
\put(295,-70){$d)$}
\end{picture}
\vskip 3.8cm

Figure 7

\end{figure}

\end{center}

\end{document}